\documentclass[onecolumn,superscriptaddress,nofootinbib,prd,preprintnumbers,amssymb,floatfix]{revtex4}

\usepackage[pdftex]{graphicx}
\usepackage{dcolumn}
\usepackage{bm}
\usepackage{graphics}
\usepackage{rotating}
\usepackage{comment}
\usepackage{multirow}
\usepackage{footnote}
\usepackage{comment}
\usepackage{epsfig}
\usepackage{color}
\usepackage{amsmath}
\usepackage{hyperref}

\newcommand{\VBFNLO}{\textsc{VBFNLO}}
\providecommand{\href}[2]{#2}

\newcommand\POWHEG{{\tt POWHEG}}
\newcommand\POWHEGBOX{{\tt POWHEG~BOX}}
\newcommand\PYTHIA{{\tt PYTHIA}}

\newcommand\HERWIG{{\tt HERWIG}}

\begin{document}

\title{Monte Carlo tools for studies of non-standard electroweak gauge boson interactions in multi-boson processes: A Snowmass White Paper}

\author{Celine Degrande}\thanks{cdegande@illinois.edu}\affiliation{Department of Physics, University of Illinois at Urbana Champaign, U.S.A}
\author{Oscar Eboli}\thanks{oeboli@gmail.com}\affiliation{Instituto de Fisica, Universidade de Sao Paulo, Sao Paulo - SP, Brazil}
\author{Bastian Feigl}\thanks{bastian.feigl@kit.edu}\affiliation{Institute for Theoretical Physics, Karlsruhe Institute of Technology (KIT), Germany}
\author{Barbara J\"ager}\thanks{jaegerba@uni-mainz.de}\affiliation{PRISMA Cluster of Excellence, Institut f\"ur Physik, Johannes-Gutenberg-Universit\"at, Mainz, Germany}
\author{Wolfgang Kilian}\thanks{kilian@physik.uni-siegen.de}\affiliation{Institut f\"ur Theoretische Physik I, Universit\"at Siegen, Germany}
\author{Olivier Mattelaer}\thanks{omatt@illinois.edu}\affiliation{Department of Physics, University of Illinois at Urbana Champaign, U.S.A. and \\
Center for Cosmology, Particle Physics and
Phenomenology (CP3), Universite Catholique de Louvain, Belgium}
\author{Michael Rauch}\thanks{michael.rauch@kit.edu}\affiliation{Institute for Theoretcial Physics, Karlsruhe Institute of
Technology (KIT), Germany}
\author{J\"urgen Reuter}\thanks{juergen.reuter@desy.de}\affiliation{DESY Theory Group, Hamburg, Germany}
\author{Marco Sekulla}\thanks{sekulla@physik.uni-siegen.de}\affiliation{Institut f\"ur Theoretische Physik I, Universit\"at Siegen, Germany}
\author{Doreen Wackeroth (ed.)}\thanks{dow@ubpheno.physics.buffalo.edu}\affiliation{Department of Physics, University at Buffalo, The State University of New York, U.S.A.}

\begin{abstract}
\vspace{50 pt}
\begin{center}
{\bf Abstract}
\end{center}
In this Snowmass 2013 white paper, we review the effective field
theory approach for studies of non-standard electroweak interactions
in electroweak vector boson pair and triple production and vector boson
scattering. We present an overview of the implementation of dimension
six and eight operators in {\textsc MadGraph5}, \VBFNLO{}, and
WHIZARD, and provide relations between the coefficients of these
higher dimensions operators used in these programs and in the
anomalous couplings approach.  We perform a tuned comparison of
predictions for multi-boson processes including non-standard
electroweak interactions with {\textsc MadGraph5}, \VBFNLO{}, and
WHIZARD. We discuss the role of higher-order corrections in these
predictions using \VBFNLO{} and a \POWHEGBOX{} implementation of
higher-order QCD corrections to $WWjj$ production. The purpose of this
white paper is to collect useful tools for the study of non-standard
EW physics at the LHC, compare them, and study the main physics issues
in the relevant processes.
\end{abstract}

\maketitle

\newpage
\tableofcontents

\section{Introduction}
\label{sec:intro}

After the LHC experiments have discovered a bosonic particle that is
fully compatible with the Standard Model (SM) Higgs boson at the level
of the electroweak (EW) precision observables, still the microscopic
mechanism of EW symmetry breaking needs to be resolved. To prove that
the SM is really the valid theory up to very high energy scales, one
either needs to overconstrain the EW sector and test its structure at
the level of next-to-leading order (NLO) corrections or find direct
evidence for a possible dynamic explanation of the Higgs
mechanism. One important ingredient is the structure of the
selfinteractions of the Higgs field, which might give a hint on its
underlying structure. For the scattering of a physical Higgs particle,
measurements of these couplings are notoriously difficult, and while a
measurement of the triple Higgs coupling seems feasible, the quartic
coupling is hopeless. However, the Higgs field also contains the
Goldstone bosons, i.e. the longitudinal modes of the EW gauge
bosons. The scattering of the longitudinal modes is overlaid with the
corresponding scattering of transversal EW gauge bosons from the
non-Abelian structure of the EW gauge group. Phenomenologically, it is
quite difficult to discriminate between them. Quartic interactions of
EW gauge bosons can be studied in either triple boson production or
vector boson scattering. There are two distinct cases: (i) where new
physics can be directly probed, and (ii) where only indirect effects
of new physics manifest themselves in the energy reach of the LHC (or
its possible energy upgrade). There are many different models fitting
the scenario (i) that are discussed in the BSM Snowmass white
paper~\footnote{See www.snowmass2013.org for the report of the
  Snowmass 2013 working group {\em The Path Beyond the Standard
    Model}.}. In all these cases, the assumption is that masses of
those resonances are approximately in the 1-5 TeV range such that they
can be directly probed at a 14 TeV machine. To cover this case in a
mostly model-independent way, simplified models have been
defined~\cite{Reuter:2013gla}, where the issue of the unitarity of the
longitudinal scattering amplitudes is carefully taken into account. In
this document, the focus is on the more pessimistic scenario that new
physics in the EW sector is out of the direct reach of the LHC or
maybe even higher energy colliders. In that case one could integrate
out new particles or resonances and one ends up with an effective
field theory (EFT) with the SM as low-energy limit. While the
translation between simplified new physics models in the EW sector to
such an EFT are described in~\cite{Reuter:2013gla}, there are also
ambiguities for the low-energy EFT. This results from the choice of
operator bases. In Section~\ref{sec:param} the EFTs in different
operator bases are discussed, and translations from one basis to
another are defined. This should simplify the comparison between many
different studies that have been performed for several past, present
and future collider experiments. In Section~\ref{sec:pred} we collect
the predictions from several studies for triboson production and
vector boson scattering at the LHC, a 33 TeV energy upgrade as well as
a 100 TeV high-energy hadron collider. In this section, also the codes
used for these predictions are introduced and described in detail. One
major purpose of this white paper is to collect useful tools for EW
physics at the LHC, compare them and study the main physics issues in
the relevant processes discussed above. These topics are described in
Section~\ref{sec:comp}. Finally, in Section~\ref{sec:summary} we
summarize our findings.

\section{Non-standard electroweak interactions}
\label{sec:param}

\subsection{Effective field theory}
\label{sec:eft}

If the energy scale of new physics is well above the energies reached in
an experiment, the new degrees of freedom cannot be produced directly and
new physics appears only as new interactions between the known
particles. These new interactions are included in the Lagrangian as
higher dimensional operators, which are invariant under the SM symmetries and
suppressed by the new physics scale $\Lambda$,
\begin{equation}
{\cal L}_{EFT} = {\cal L}_{SM} + \sum_{d>4}\sum_i \frac{\tilde c_i}{\Lambda^{d-4}}{\cal O}_i
\label{eq:eft}
\end{equation}
where $d$ is the dimension of the operators. In the limit
$\Lambda\to\infty$, this Lagrangian reduces to the SM one. Since the
coefficients of the higher dimensional operators, $\tilde c_i$, are
fixed by the complete high-energy theory, any extension of the SM can
be parametrized by this Lagrangian, where the $\tilde c_i$ are kept as
free parameters. Below the new physics scale, only the operators with
lowest dimensions can give a large contribution and should therefore
be kept. In particular, the SM contribution is expected to be larger
than the new physics one. Once truncated, the Lagrangian becomes
predictive even without fixing the coefficients and parametrizes any
heavy new physics scenario. However, it should be kept in mind that
this truncated Lagrangian is only valid below the new physics scale.

In the following, we will describe the EFT of new physics including dimension six and dimension eight operators
that modify the interactions among electroweak gauge bosons:
\begin{equation}
{\cal L}_{EFT} = {\cal L}_{SM} + \sum_{i=WWW,W,B, \\\Phi W,\Phi B} \frac{c_i}{\Lambda^{2}}{\cal O}_i+\sum_{j=1,2} 
\frac{f_{S,j}}{\Lambda^{4}}{\cal O}_{S,j}+\sum_{j=0,\ldots,9}  \frac{f_{T,j}}{\Lambda^{4}}{\cal O}_{T,j}+\sum_{j=0,\ldots,7}\frac{f_{M,j}}{\Lambda^{4}}{\cal O}_{M,j}
\label{eq:eft2}
\end{equation}

\subsection{Dimension-six operators for electroweak vector boson pair and triple production and scattering}
\label{sec:dim6}

If baryon and lepton numbers are conserved, only operators with even
dimension can be constructed. Consequently, the largest new physics
contribution is expected from dimension-six operators.  Three CP
conserving dimension-six operators,
\begin{equation}
 \begin{aligned}
   {\cal O}_{WWW}&=\mbox{Tr}[W_{\mu\nu}W^{\nu\rho}W_{\rho}^{\mu}]\\
{\cal O}_W&=(D_\mu\Phi)^\dagger W^{\mu\nu}(D_\nu\Phi)\\
{\cal O}_B&=(D_\mu\Phi)^\dagger B^{\mu\nu}(D_\nu\Phi),
\end{aligned}
\label{opTGC}
\end{equation}

and two CP violating dimension-six operators,
\begin{equation}
 \begin{aligned}
{\cal O}_{\tilde WWW}&=\mbox{Tr}[{\tilde W}_{\mu\nu}W^{\nu\rho}W_{\rho}^{\mu}]\\
{\cal O}_{\tilde W}&=(D_\mu\Phi)^\dagger {\tilde W}^{\mu\nu}(D_\nu\Phi),
\end{aligned}
\label{opTGCCP}
\end{equation}
affect the triple and quartic gauge couplings. Here $\Phi$ denotes the Higgs doublet field and
the covariant derivative for such a field with hypercharge $Y=1/2$ is given by
\begin{equation}
 D_\mu \equiv \partial_\mu + i \frac{g'}{2} B_\mu  + i g W_\mu^i \frac{\tau^i}{2} 
\label{eq:def_dmu}
\end{equation}
where $\tau^i, i=1,2,3$ are the $SU(2)_I$ generators  
with $\mbox{Tr}[ \tau^i \tau^j ] = 2\delta^{ij}$. 
The field strength tensors of the $SU(2)_I$ ($W^i_\mu$) and $U(1)_Y$ ($B_\mu$)  gauge fields read
\begin{equation}
 \begin{aligned}
W_{\mu\nu} & =  \frac{i}{2} g\tau^i (\partial_\mu W^i_\nu - \partial_\nu W^i_\mu
	+ g \epsilon_{ijk} W^j_\mu W^k_\nu ) \\ 
B_{\mu \nu} & = \frac{i}{2} g' (\partial_\mu B_\nu - \partial_\nu B_\mu)  \; .
\end{aligned}
\label{eq:fields}
\end{equation}
Like in the SM, TGCs and
QGCs from dimension-six operators are completely related to guarantee
gauge invariance.  In addition, three CP-conserving 
operators
\begin{equation}
 \begin{aligned}
\mathcal{O}_{\Phi d} &= \partial_\mu\left(\Phi^\dagger \Phi\right)\partial^\mu\left(\Phi^\dagger \Phi\right)\\
\mathcal{O}_{\Phi W} &= \left(\Phi^\dagger\Phi\right) \mbox{Tr}[W^{\mu\nu}W_{\mu\nu}] \\
\mathcal{O}_{\Phi B} &= \left(\Phi^\dagger\Phi\right)  B^{\mu\nu}B_{\mu\nu}  
\end{aligned}
\label{opphi} 
\end{equation}
and two CP-violating dimension-six operators
\begin{equation}
 \begin{aligned}
\mathcal{O}_{\tilde{W}W} &= \Phi^{\dagger} {\tilde{W}}_{\mu\nu} {W}^{\mu\nu} \Phi \\
\mathcal{O}_{\tilde{B}B} &= \Phi^{\dagger} {\tilde{B}}_{\mu\nu} {B}^{\mu\nu} \Phi 
\end{aligned}
\label{eq:cpodd} 
\end{equation}
modify the coupling of the Higgs to the weak gauge bosons and therefore
the four-gauge-boson amplitudes.
The list of vertices relevant to three- and four-gauge-boson amplitudes 
of each operator is displayed in Tab.~\ref{tab:vert}.
\begin{table}[ht]
\begin{tabular}{l|c|c|c|c|c|c|c|c|c|c} \hline
   					&ZWW	&AWW	&HWW	&HZZ	&HZA	&HAA	&WWWW	&ZZWW	&ZAWW	&AAWW	\\
   \hline
   ${\cal O}_{WWW}$		&X		&X		&		&		&		&		&X		&X		&X		&X\\
   ${\cal O}_{W}$		&X		&X		&X		&X		&X		&		&X		&X		&X		&\\
   ${\cal O}_{B}$		&X		&X		&		&X		&X		&		&		&		&		&\\
   ${\cal O}_{\Phi d}$		&		&		&X		&X		&		&		&		&		&		&\\
   ${\cal O}_{\Phi W}$		&		&		&X		&X		&X		&X		&		&		&		&\\
   ${\cal O}_{\Phi B}$ 		&		&		&		&X		&X		&X		&		&		&		&\\
   \hline
   ${\cal O}_{\tilde WWW}$	&X		&X		&		&		&		&		&X		&X		&X		&X\\
   ${\cal O}_{\tilde W}$	&X		&X		&X		&X		&X		&		&		&		&		& \\
${\cal O}_{\tilde WW}$	&		&		&X		&X		&X		&X		&		&		&		&\\
${\cal O}_{\tilde BB}$	&		&		&		&X		&X		&X		&		&		&		& \\ \hline
\end{tabular}
\caption{The vertices induced by each operator are marked with X in the corresponding column. The vertices that are not relevant for three- and four-gauge-boson amplitudes have been omitted.}
\label{tab:vert}
\end{table}
We have neglected the operators affecting the couplings of the bosons
to fermions as they can be measured in other processes such as $Z$
decay.  This is a minimal set of independent dimension-six operators
relevant to 
amplitudes involving vertices of three and four electroweak gauge bosons. 
Additional dimension-six operators invariant under SM
symmetries can be constructed but they can be shown to be equivalent to a linear
combination of the previous operators by using equations of
motion. Consequently, the choice of basis of operators is not unique and other choices than the one presented here 
can be found in the literature. For
example, the operators $Q_{\Phi D}$ and $Q_{\Phi WB}$ in
Ref.~\cite{Grzadkowski:2010es} have been replaced in this paper by
${\cal O}_W$ and ${\cal O}_B$. Our basis avoids the otherwise
necessary redefinition of the masses of the gauge bosons and the
mixing of the neutral vector bosons.  The operator $\mathcal{O}_{\Phi
  d}$ does not contain any gauge boson since $\Phi^\dagger \Phi$ is a
singlet under all the SM gauge groups. However, it contributes to the
Higgs field's kinetic term after $\Phi$ has been replaced by its value
in the unitary gauge, i.e. with
\begin{equation}
\Phi=\left(0,\frac{v+h}{\sqrt2}\right)^T
\label{eq:unitary}
\end{equation}
one finds
\begin{equation}
\mathcal{O}_{\Phi d} \ni v^2 \partial_\mu h \partial^\mu h,
\end{equation}
and it requires a renormalization of the Higgs field,
\begin{equation}
h \to h(1-\frac{c_{\Phi d}}{\Lambda^2} v^2),
\end{equation}
in the full Lagrangian. The Higgs couplings to all particles
including the electroweak gauge bosons are consequently multiplied by
the same factor. $\mathcal{O}_{\Phi W}$ and $\mathcal{O}_{\Phi B}$
modify the kinetic term of the gauge bosons after the Higgs doublet
has been replaced by its vacuum expectation value ($v$). Those two operators require then a
renormalization of the gauge fields and the gauge couplings. As a
matter of fact, their part proportional to $v^2$ is entirely absorbed
by those redefinitions and can therefore be removed directly in the
definition of the operators, i.e.
\begin{equation}
 \begin{aligned}
  \mathcal{O}_{\Phi W} &= \left(\Phi^\dagger\Phi-v^2\right) \mbox{Tr}[W^{\mu\nu}W_{\mu\nu}] \\
\mathcal{O}_{\Phi B} &= \left(\Phi^\dagger\Phi-v^2\right)  B^{\mu\nu}B_{\mu\nu}  
 \end{aligned}
\end{equation}
It is now clear that those operators affect only the vertices with one
or two Higgs boson and not the TGCs or the QGCs.

\subsection{Dimension-eight operators for genuine QGCs}
\label{sec:dim8}

As can be seen in Table~\ref{tab:vert}, the dimension--six operators giving
rise to QGCs also exhibit TGCs. In order to separate the effects of
the QGCs we shall consider effective operators that lead to QGCs
without a TGC associated to them. Moreover, not all possible QGCs are
generated by dimension--six operators, for instance, these operators
do not give rise to quartic couplings among the neutral gauge bosons
\footnote{Notice that the lowest order operators leading to neutral
  TGCs are also of dimension eight. }.  The lowest dimension operator
that leads to quartic interactions but does not exhibit two or three
weak gauge boson vertices is of dimension eight\footnote{Effective
  operators possessing QCGs but no TGCs can be generated at tree level by
  new physics at a higher scale~\cite{Arzt:1994gp}, in contrast with
  operators containing TGCs that are generated at loop level.  }. The
counting is straight forward: we can get a weak boson field either
from the covariant derivative ($D_\mu$ of Eq.~\ref{eq:def_dmu}) of $\Phi$ 
or from the field strength
tensor of Eq.~\ref{eq:fields}. In either case, the vector field is accompanied by $v$ (after using Eq.~\ref{eq:unitary}) 
or a derivative $\partial_\mu$. Therefore, genuine quartic vertices are of dimension 8 or
higher.

The idea behind using dimension--eight operators for QGCs is that the
anomalous QGCs are to be considered as a straw man to evaluate the LHC
potential to study these couplings, without having any theoretical
prejudice about their size.  There are three classes of genuine QGC
operators~\cite{Eboli:2006wa}:

\subsubsection{Operators containing only  $D_\mu\Phi$}

This class contains two independent operators, {\em i.e.}
\begin{align}
  {\cal O}_{S,0} &= \left [ \left ( D_\mu \Phi \right)^\dagger
 D_\nu \Phi \right ] \times
\left [ \left ( D^\mu \Phi \right)^\dagger
D^\nu \Phi \right ] \; ,
\\
  {\cal O}_{S,1} &= \left [ \left ( D_\mu \Phi \right)^\dagger
 D^\mu \Phi  \right ] \times
\left [ \left ( D_\nu \Phi \right)^\dagger
D^\nu \Phi \right ] \; ,
\end{align}
where 
the
Higgs covariant derivative is given by the expression  in
Eq.~\ref{eq:def_dmu}. These operators can be generated when we integrate
out a spin--one resonance that couples to gauge--boson pairs with
\begin{equation}
\frac{f_{S,0}}{\Lambda^4} = - \frac{f_{S,1}}{\Lambda^4} = \frac{12 \pi}{M_\rho^4}
~\frac{\Gamma_\rho}{M_\rho} \; ,
\end{equation}
where $M_\rho$ ($\Gamma_\rho$) is the mass (width) of the vector resonance
~\cite{Bagger:1992vu}.

The operators ${\cal O}_{S,0}$ and ${\cal O}_{S,1}$ contain quartic
$W^+W^-W^+W^-$, $W^+W^-ZZ$ and $ZZZZ$ interactions that do not depend
on the gauge boson momenta; for a comparative table showing all QGCs
induced by dimension--eight operators see Table~\ref{tab:vertices}.
In our framework, the QGCs are accompanied by vertices with more than
4 particles due to gauge invariance.  In order to simply rescale the
SM quartic couplings containing $W^\pm$ and $Z$ it is enough to choose
$f_{S,0} = - f_{S,1} = f$ which leads to SM quartic couplings modified by
a factor $(1+f v^4/8)$, where $v$ is the Higgs vacuum expectation
value ($v \simeq 246$~GeV).

\subsubsection{ Operators containing $D_\mu\Phi$ and two field strength tensors}

QGCs are also generated by considering two electroweak field strength
tensors and two covariant derivatives of the Higgs
doublet~\cite{Eboli:2006wa}:
%
\begin{align}
 {\cal O}_{M,0} &=   \hbox{Tr}\left [ {W}_{\mu\nu} {W}^{\mu\nu} \right ]
\times  \left [ \left ( D_\beta \Phi \right)^\dagger
D^\beta \Phi \right ] \; ,
\\
 {\cal O}_{M,1} &=   \hbox{Tr}\left [ {W}_{\mu\nu} {W}^{\nu\beta} \right ]
\times  \left [ \left ( D_\beta \Phi \right)^\dagger
D^\mu \Phi \right ] \; ,
\\
 {\cal O}_{M,2} &=   \left [ B_{\mu\nu} B^{\mu\nu} \right ]
\times  \left [ \left ( D_\beta \Phi \right)^\dagger
D^\beta \Phi \right ] \; ,
\\
 {\cal O}_{M,3} &=   \left [ B_{\mu\nu} B^{\nu\beta} \right ]
\times  \left [ \left ( D_\beta \Phi \right)^\dagger
D^\mu \Phi \right ] \; ,
\\
  {\cal O}_{M,4} &= \left [ \left ( D_\mu \Phi \right)^\dagger {W}_{\beta\nu}
 D^\mu \Phi  \right ] \times B^{\beta\nu} \; ,
\\
  {\cal O}_{M,5} &= \left [ \left ( D_\mu \Phi \right)^\dagger {W}_{\beta\nu}
 D^\nu \Phi  \right ] \times B^{\beta\mu} \; ,
\\
  {\cal O}_{M,6} &= \left [ \left ( D_\mu \Phi \right)^\dagger {W}_{\beta\nu}
{W}^{\beta\nu} D^\mu \Phi  \right ] \; ,
\\
  {\cal O}_{M,7} &= \left [ \left ( D_\mu \Phi \right)^\dagger {W}_{\beta\nu}
{W}^{\beta\mu} D^\nu \Phi  \right ] \; ,
\end{align}
where the field strengths $W_{\mu\nu}$ and $B_{\mu\nu}$ have been
defined above in Eq.~\eqref{eq:fields}. In this class of effective operators the
quartic gauge-boson interactions depend upon the momenta of the vector
bosons due to the presence of the field strength in their
definitions. Therefore, the Lorentz structure of these operators can
not be reduced to the SM one. The complete list of quartic vertices
modified by these operators can be found in Table~\ref{tab:vertices}.

\subsubsection{Operators containing only field strength tensors}

The following operators containing four field strength tensors
also lead to quartic anomalous couplings:
%
\begin{align}
 {\cal O}_{T,0} &=   \hbox{Tr}\left [ {W}_{\mu\nu} {W}^{\mu\nu} \right ]
\times   \hbox{Tr}\left [ {W}_{\alpha\beta} {W}^{\alpha\beta} \right ] \; ,
\\
 {\cal O}_{T,1} &=   \hbox{Tr}\left [ {W}_{\alpha\nu} {W}^{\mu\beta} \right ] 
\times   \hbox{Tr}\left [ {W}_{\mu\beta} {W}^{\alpha\nu} \right ] \; ,
\\
 {\cal O}_{T,2} &=   \hbox{Tr}\left [ {W}_{\alpha\mu} {W}^{\mu\beta} \right ]
\times   \hbox{Tr}\left [ {W}_{\beta\nu} {W}^{\nu\alpha} \right ]  \; ,
\\
%
%
 {\cal O}_{T,5} &=   \hbox{Tr}\left [ {W}_{\mu\nu} {W}^{\mu\nu} \right ]
\times   B_{\alpha\beta} B^{\alpha\beta}  \; ,
\\
 {\cal O}_{T,6} &=   \hbox{Tr}\left [ {W}_{\alpha\nu} {W}^{\mu\beta} \right ]
\times   B_{\mu\beta} B^{\alpha\nu}  \; ,
\\
 {\cal O}_{T,7} &=   \hbox{Tr}\left [ {W}_{\alpha\mu} {W}^{\mu\beta} \right ]
\times   B_{\beta\nu} B^{\nu\alpha}  \; ,
\\
 {\cal O}_{T,8} &=   B_{\mu\nu} B^{\mu\nu}  B_{\alpha\beta} B^{\alpha\beta}
\\
 {\cal O}_{T,9} &=  B_{\alpha\mu} B^{\mu\beta}   B_{\beta\nu} B^{\nu\alpha}  \; .
\end{align}
It is interesting to note that the two last operators $ {\cal
  O}_{T,8}$ and $ {\cal O}_{T,9}$ give rise to QGCs containing only
the neutral electroweak gauge bosons.

Previous analyses~\cite{Belyaev:1998ih,Eboli:2000ad,Eboli:2003nq} of the LHC potential to study
QGCs were based on the non--linear realization of the gauge symmetry,
{\em i.e.} using chiral Lagrangians as for instance implemented in WHIZARD. 
The relation between the above framework
and chiral Lagrangians can be found in Section~\ref{sec:whizard-dim8}.

\begin{table}[h]
\begin{tabular}{|c|c|c|c|c|c|c|c|c|c|} \hline
   & WWWW & WWZZ & ZZZZ & WWAZ& WWAA& ZZZA & ZZAA& ZAAA& AAAA\\
\hline
${\cal O}_{S,0}$, ${\cal O}_{S,1}$ &  X & X & X &   &   &   &   &   &  \\
\hline 
${\cal O}_{M,0}$, ${\cal O}_{M,1}$,${\cal O}_{M,6}$ ,${\cal O}_{M,7}$ 
 &  X & X & X & X & X & X & X &   &   \\
\hline 
${\cal O}_{M,2}$ ,${\cal O}_{M,3}$, 
${\cal O}_{M,4}$ ,${\cal O}_{M,5}$ 
&    & X & X & X & X & X & X &   &   \\ \hline
${\cal O}_{T,0}$ ,${\cal O}_{T,1}$ ,${\cal O}_{T,2}$ 
& X   & X & X & X & X & X & X & X  &X   \\ \hline
${\cal O}_{T,5}$ ,${\cal O}_{T,6}$ ,${\cal O}_{T,7}$ 
&    & X & X & X & X & X & X & X  &X   \\ \hline
${\cal O}_{T,8}$ ,${\cal O}_{T,9}$ 
&    &  & X &  &  & X & X & X  &X   \\ \hline
\end{tabular}
\caption{Quartic vertices modified by each dimension-8 
operator are marked with $X$.}
\label{tab:vertices}
\end{table}

\subsection{Comparison with the anomalous coupling approach and the LEP convention for aQGCs}
\label{sec:anomalous}

The anomalous couplings approach is based on the Lagrangian
\cite{Hagiwara:1986vm}
\begin{equation}
 \begin{aligned}
  {\cal L}=&ig_{WWV}\left(g_1^V(W_{\mu\nu}^+W^{-\mu}-W^{+\mu}W_{\mu\nu}^-)V^\nu
+\kappa_VW_\mu^+W_\nu^-V^{\mu\nu}
+\frac{\lambda_V}{M_W^2}W_\mu^{\nu+}W_\nu^{-\rho}V_\rho^{\mu}
\right.\\&\left.
+ig_4^VW_\mu^+W^-_\nu(\partial^\mu V^\nu+\partial^\nu V^\mu)
-ig_5^V\epsilon^{\mu\nu\rho\sigma}(W_\mu^+\partial_\rho W^-_\nu-\partial_\rho W_\mu^+W^-_\nu)V_\sigma
\right.\\&\left.
+\tilde{\kappa}_VW_\mu^+W_\nu^-\tilde{V}^{\mu\nu}
+\frac{\tilde{\lambda}_V}{m_W^2}W_\mu^{\nu+}W_\nu^{-\rho}\tilde{V}_\rho^{\mu}
\right) \, ,
 \end{aligned}
 \label{eq:L}
\end{equation}
where $V=\gamma,Z$; $W_{\mu\nu}^\pm = \partial_\mu W_\nu^\pm -
\partial_\nu W_\mu^\pm$, $V_{\mu\nu} = \partial_\mu V_\nu -
\partial_\nu V_\mu$, $g_{WW\gamma}=-e$ and $g_{WWZ}=-e\cot\theta_W$.
The first three terms of Eq.~\ref{eq:L} are $C$ and $P$ invariant
while the remaining four terms violate $C$ and/or $P$.
Electromagnetic gauge invariance requires that $g_1^\gamma =1$ and
$g_4^\gamma=g_5^\gamma = 0$.  Finally there are five independent $C$-
and $P$-conserving parameters: $g_1^Z, \kappa_\gamma, \kappa_Z,
\lambda_\gamma, \lambda_Z$; and six $C$ and/or $P$ violating
parameters: $g_4^Z, g_5^Z, \tilde{\kappa}_\gamma, \tilde{\kappa_Z},
\tilde{\lambda}_\gamma, \tilde{\lambda_Z}$. This Lagrangian is not the
most generic one as extra derivatives can be added in all the
operators. Furthermore, there is no reason to remove those extra terms
since they are not suppressed by $\Lambda$ but by $M_W$.

The effective field theory approach described in the previous section
allows one to calculate those parameters in terms of the coefficients
of the five dimension-six operators relevant for TGCs, i.e. in terms of the EFT 
coefficients $c_{WWW}, c_W, c_B, c_{\tilde{W}WW}$ and $ c_{\tilde{W}}$.
One finds for the anomalous TGC parameters\cite{Hagiwara:1993ck,Wudka:1994ny}:
\begin{align}
g_1^Z &= 1+c_W\frac{m_Z^2}{2\Lambda^2}\\
\kappa_\gamma &= 1+(c_W+c_B)\frac{m_W^2}{2\Lambda^2}\\
\kappa_Z &= 1+(c_W-c_B\tan^2\theta_W)\frac{m_W^2}{2\Lambda^2}\\
\lambda_\gamma &= \lambda_Z = c_{WWW}\frac{3g^2m_W^2}{2\Lambda^2}\\
g_4^V &= g_5^V=0\\
\tilde{\kappa}_\gamma &=
c_{\tilde{W}}\frac{m_W^2}{2\Lambda^2}\\
\tilde{\kappa}_Z &=
-c_{\tilde{W}}\tan^2\theta_W\frac{m_W^2}{2\Lambda^2}\\
\tilde{\lambda}_\gamma &= \tilde{\lambda}_Z = c_{\tilde{W}WW}\frac{3g^2m_W^2}{2\Lambda^2}
\end{align}
Defining $\Delta g_1^Z = g_1^Z - 1$, $\Delta \kappa_{\gamma,Z} = \kappa_{\gamma,Z} - 1$, the relation \cite{Hagiwara:1993ck}
\begin{equation}
\Delta g_1^Z=\Delta \kappa_Z + \tan^2\theta_W \Delta \kappa_\gamma
\end{equation}
and the relation $\lambda_\gamma = \lambda_Z$ reduce the five $C$ and $P$ conserving parameters down to three.  For the $C$ and/or $P$ violating parameters, the relation
\begin{equation}
0=\tilde \kappa_Z + \tan^2\theta_W \tilde \kappa_\gamma
\end{equation}
and the relations $\tilde\lambda_\gamma = \tilde\lambda_Z$ and
$g_4^Z=g_5^Z=0$ reduce the six $C$ and/or $P$ violating parameters
down to just two.

The Lagrangian of Eq.~\ref{eq:L} is not $SU(2)_L$ gauge invariant even
after imposing those relations because the quartic and higher
multiplicity couplings are not included. Furthermore, gauge invariance
requires also several relations between vertices with different number
of particles. Therefore, the anomalous coupling Lagrangian cannot be
used for four-gauge-boson amplitudes.
The quartic couplings involving two photons have been parametrized in
a similar way. However, the parametrization is not generic enough and
does not include the contributions from the dimension-six operators.

The LEP2 constraints on the vertices $\gamma\gamma W^+ W^-$ and
$\gamma\gamma ZZ$~\cite{quarticATlep} described in terms of anomalous couplings
$a_0/\Lambda^2$ and $a_c/\Lambda^2$ can be translated into bounds on
%
%
%
%
%
%
%
%
%
%
$f_{M,0}$ -- $f_{M,7}$. 
%
%
%
In Ref.~\cite{Stirling:1999ek} (see also Refs~\cite{Belanger:1992qh,Eboli:1993wg}), genuine
anomalous quartic couplings involving two photons have been introduced as follows:
\begin{equation}
\begin{aligned}
{\cal L}_0 & =  - \frac{e^2}{16 \pi \Lambda^2} a_0 F_{\mu \nu} F^{\mu \nu} \vec{W}^\alpha \vec{W}_\alpha \\
{\cal L}_c & =  - \frac{e^2}{16 \pi \Lambda^2} a_c F_{\mu \alpha} F^{\mu \beta} \vec{W}^\alpha \vec{W}_\beta
\end{aligned}
\label{eq:aoac}
\end{equation}
with
\begin{equation}
\begin{aligned}
F^{\mu \nu} & =\partial^\mu A^\nu - \partial^\nu A^\mu  \\
\vec{W}_{\mu} & = 
\left(
\begin{array}{c}
\frac{1}{\sqrt{2}} (W_\mu^+ + W_\mu^-) \\
\frac{i}{\sqrt{2}} (W_\mu^+ - W_\mu^-) \\
\frac{Z_\mu}{\cos\theta_w} 
\end{array}
\right)
\end{aligned}
\label{eq:LEPfields}
\end{equation}
where $A_\mu$ and $W_\mu^\pm,Z_\mu$ denote the photon and weak fields, respectively.
%
%
%
%
%
%
%
%
%
%
Thus, using the conventions of Eq.~\ref{eq:fields} for the fields in the operators $\mathcal{O}_{M,i}$,
and Eq.~\ref{eq:LEPfields} for the fields in the operators ${\cal L}_0 / {\cal L}_c$, 
the following relations for the $WW\gamma\gamma$ (upper sign) and $ZZ\gamma\gamma$ (lower sign) vertices can be derived:
%
   \begin{alignat}{3}
    \frac{f_{M,0}}{\Lambda^4} &= \hspace{2ex} \frac{a_0}{\Lambda^2} \frac{1}{g^2 v^2}
    &&\qquad\text{and}\qquad&
   \frac{f_{M,1}}{\Lambda^4} &= -\frac{a_c}{\Lambda^2} \frac{1}{g^2 v^2}  \\
    \frac{f_{M,2}}{\Lambda^4} &=\hspace{2ex} \frac{a_0}{\Lambda^2} \frac{2}{g^2 v^2}
    &&\qquad\text{and}\qquad&
   \frac{f_{M,3}}{\Lambda^4} &= -\frac{a_c}{\Lambda^2} \frac{2}{g^2 v^2}  \\
    \frac{f_{M,4}}{\Lambda^4} &= \pm\frac{a_0}{\Lambda^2} \frac{1}{g^2 v^2}
    &&\qquad\text{and}\qquad&
   \frac{f_{M,5}}{\Lambda^4} &= \pm\frac{a_c}{\Lambda^2} \frac{2}{g^2 v^2}  \\
    \frac{f_{M,6}}{\Lambda^4} &= \hspace{2ex}\frac{a_0}{\Lambda^2} \frac{2}{g^2 v^2}
    &&\qquad\text{and}\qquad&
   \frac{f_{M,7}}{\Lambda^4} &= \hspace{2ex}\frac{a_c}{\Lambda^2} \frac{2}{g^2 v^2}  \, .
  \end{alignat}

\subsection{Conventions for non-standard electroweak gauge boson interactions in different programs}
\label{sec:conventions}

\subsubsection{Dimension-8 operators: VBFNLO and MadGraph5}
\label{sec:mg-dim8}

The convention for the dimension-8-operators in VBFNLO is
the same as described in Section~\ref{sec:dim8}, and the
coefficients $f_i/{\Lambda^4}$ set in the input file are the
ones that multiply the operators of Section~\ref{sec:dim8}.
However, the MadGraph5 implementation by means of a UFO file~\cite{Eboli:anom4}
uses expressions for the field strengths which are slightly
different than the ones from Eq.~\ref{eq:fields}:
\begin{equation}
\begin{aligned}
\widehat{W}_{\mu\nu} & \;=\;  \frac{1}{2} \tau^i (\partial_\mu W^i_\nu - \partial_\nu W^i_\mu
	+ g \epsilon_{ijk} W^j_\mu W^k_\nu ) = \frac{1}{i g}  W_{\mu\nu} \\ 
\widehat{B}_{\mu \nu} & \;=\;  (\partial_\mu B_\nu - \partial_\nu B_\mu) =  \frac{2}{i g'}   B_{\mu\nu}
\end{aligned}
\end{equation}
The resulting changes can be absorbed in a redefinition of the operator coefficients:
\begin{align}
 f_{S,0,1}   &= f_{S,0,1}^\text{VBFNLO}   \;=\;                            f_{S,0,1}^\text{MG5}   \\
 f_{M,0,1}   &= f_{M,0,1}^\text{VBFNLO}   \;=\; - \frac{1}{g^2}      \cdot f_{M,0,1}^\text{MG5}   \\
 f_{M,2,3}   &= f_{M,2,3}^\text{VBFNLO}   \;=\; - \frac{4}{g'^2}     \cdot f_{M,2,3}^\text{MG5}   \\
 f_{M,4,5}   &= f_{M,4,5}^\text{VBFNLO}   \;=\; - \frac{2}{g g'}     \cdot f_{M,4,5}^\text{MG5}   \\
 f_{M,6,7}   &= f_{M,6,7}^\text{VBFNLO}   \;=\; - \frac{1}{g^2}      \cdot f_{M,6,7}^\text{MG5}   \\
 f_{T,0,1,2} &= f_{T,0,1,2}^\text{VBFNLO} \;=\;   \frac{1}{g^4}      \cdot f_{T,0,1,2}^\text{MG5} \\
 f_{T,5,6,7} &= f_{T,5,6,7}^\text{VBFNLO} \;=\;   \frac{4}{g^2 g'^2} \cdot f_{T,5,6,7}^\text{MG5} \\
 f_{T,8,9}   &= f_{T,8,9}^\text{VBFNLO}   \;=\;   \frac{16}{g'^4}    \cdot f_{T,8,9}^\text{MG5}
\end{align}

\subsubsection{Dimension-8 operators: WHIZARD}
\label{sec:whizard-dim8}

As WHIZARD uses different anomalous couplings operators than the ones described
in Section~\ref{sec:dim8}, assuming
a different symmetry group~\cite{Alboteanu:2008my}, a conversion is in general not possible.
However, a vertex-specific conversion exists for the operators ${\cal O}_{S,0}$
and ${\cal O}_{S,1}$ to their corresponding operators
\begin{align}
   {\cal L}^{(4)}_4 &= \alpha_4 \left[ \textrm{Tr} \left( V_\mu V_\nu \right) \right] ^2  \\
   {\cal L}^{(4)}_5 &= \alpha_5 \left[ \textrm{Tr} \left( V_\mu V^\mu \right) \right] ^2 , \quad \textrm{with}\; V_\mu = \left( D_\mu \Sigma \right) \Sigma^\dagger \; .
\end{align}
The conversion reads:
\begin{itemize}
 \item for the WWWW-Vertex: 
  \begin{align}
     \alpha_4 &=  \frac{f_{S,0}}{\Lambda^4} \frac{v^4}{8}\\
     \alpha_4 + 2 \cdot \alpha_5 &=  \frac{f_{S,1}}{\Lambda^4} \frac{v^4}{8}
  \end{align}
 \item for the WWZZ-Vertex:
  \begin{align}
    \alpha_4 &=  \frac{f_{S,0}}{\Lambda^4} \frac{v^4}{16}\\
    \alpha_5 &=  \frac{f_{S,1}}{\Lambda^4} \frac{v^4}{16}
  \end{align}
 \item for the ZZZZ-Vertex:
  \begin{align}
    \alpha_4 + \alpha_5 =  \left(\frac{f_{S,0}}{\Lambda^4} + \frac{f_{S,1}}{\Lambda^4} \right) \frac{v^4}{16}
  \end{align}
\end{itemize}

\subsubsection{Dimension-6 operators: VBFNLO and MadGraph5}
\label{sec:mg-dim6}

The MadGraph model EWdim6 has been generated from FeynRules and
contains the operators from Eqs.~\ref{opTGC}, \ref{opTGCCP}
and \ref{opphi}, with the exception of ${\cal O}_{\tilde WW}$, ${\cal O}_{\tilde BB}$ and 
${\cal O}_{D\tilde W}$~\footnote{We have neglected the CP
  violating operators with the dual strength tensors affecting only the gauge boson Higgs couplings, since measuring CP
  violation in the four-weak-boson amplitude would be very
  challenging.}. The names of the coefficients is displayed in
Tab.~\ref{tab:cname}.  All the coefficients include the $1/\Lambda^2$
as reminded by the "L2" at the end of the names and are in
TeV$^{-2}$. The model also has a new coupling order $NP$ counting the
power of $1/\Lambda$. Consequently, each vertex from the dimension-six
operators has NP=2.

\begin{table}[ht]
\centering
 \begin{tabular}{l|l}
 $c_{WWW}/\Lambda^2$ & CWWWL2\\\hline
 $c_{W}/\Lambda^2$ & CWL2\\\hline
 $c_{B}/\Lambda^2$ & CBL2\\\hline
 $c_{\tilde{W}WW}/\Lambda^2$ & CPWWWL2\\\hline
 $c_{\tilde{W}}/\Lambda^2$ & CPWL2\\\hline
 $c_{\Phi d}/\Lambda^2$ & CphidL2\\\hline
 $c_{\Phi W}/\Lambda^2$ & CphiWL2\\\hline
 $c_{\Phi B}/\Lambda^2$ & CphiBL2\\
 \end{tabular}
 \caption{Names of the couplings of the dimension-six operators present in the EWdim6 model of MadGraph5.}
 \label{tab:cname}
\end{table}


The operators from Eqs.~\ref{opTGC} and \ref{opTGCCP} in Section~\ref{sec:dim6}
are directly available in VBFNLO. From Eq.~\ref{opphi} the operators $\mathcal{O}_{\tilde{W}W}$,
$\mathcal{O}_{\tilde{B}B}$ and $\mathcal{O}_{\Phi B}$ are available as well (${\cal O}_{\Phi B}$ is 
called ${\cal O}_{BB}$ within VBFNLO).
Additionally, the operator 
\begin{equation}
 \mathcal{O}_{WW} = \Phi^{\dagger} {W}_{\mu\nu} {W}^{\mu\nu} \Phi
\end{equation}
from VBFNLO can be related to the operator
${\cal O}_{\Phi W}$ by choosing the coefficient as
\begin{align}
 c_{WW} &= 2 \cdot c_{\Phi W}
\end{align}
In addition to those operators, VBFNLO also provides the following CP-odd operators:
\begin{equation}
 \begin{aligned}
\mathcal{O}_{\tilde{B}} &= (D_{\mu} \Phi)^{\dagger} {\tilde{B}}^{\mu \nu} (D_{\nu} \Phi)  \\
\mathcal{O}_{B\tilde{W}} &= \Phi^{\dagger} {B}_{\mu\nu} {\tilde{W}}^{\mu\nu} \Phi  \\
\mathcal{O}_{D\tilde{W}} &= \textrm{Tr} \left( [D_\mu, {\tilde{W}}_{\nu\rho} ] [D^\mu, {\tilde{W}}^{\nu\rho}] \right) \, .
\end{aligned}
\end{equation}
However, only 4 of the 7 CP-odd operators are linearly independent, so 
the additional operators can be expressed in terms of the operators of Eqs.~\ref{opTGCCP} and \ref{eq:cpodd} as follows:
\begin{equation}
 \begin{aligned}
\mathcal{O}_{\tilde{B}} &= \mathcal{O}_{\tilde{W}} + \frac{1}{2} \mathcal{O}_{\tilde{W}W} - \frac{1}{2} \mathcal{O}_{\tilde{B}B}   \\
\mathcal{O}_{B\tilde{W}} &= -2 \, \mathcal{O}_{\tilde{W}} - \mathcal{O}_{\tilde{W}W} \\\
\mathcal{O}_{D\tilde{W}} &= -4 \, \mathcal{O}_{\tilde{W}WW} \, . 
 \end{aligned}
\end{equation}
The CP-conserving anomalous couplings implementation is also available in  VBFNLO with the
parameters $\Delta g_1^Z$, $\Delta \kappa_Z$, $\Delta \kappa_\gamma$, and $\lambda_\gamma$,
defined in Section~\ref{sec:anomalous}.

\subsection{Discussion of unitarity bounds and usage of form factors}
\label{sec:unitarity}

\begin{figure}[h]
\centering
\includegraphics[width=0.48\textwidth]{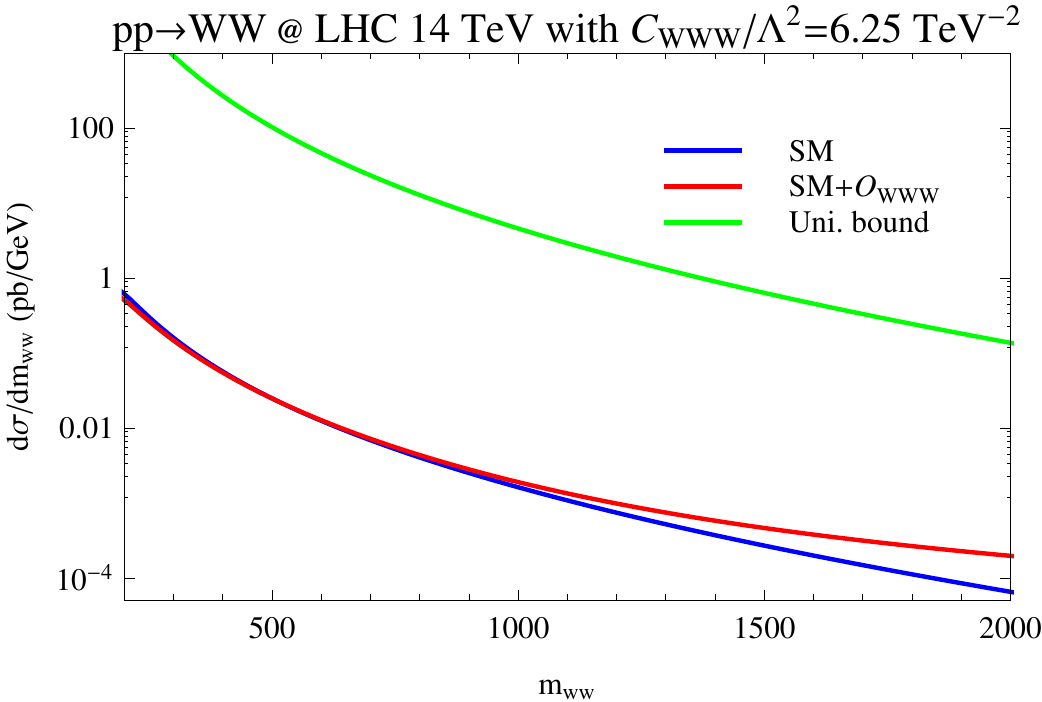}\includegraphics[width=0.48\textwidth]{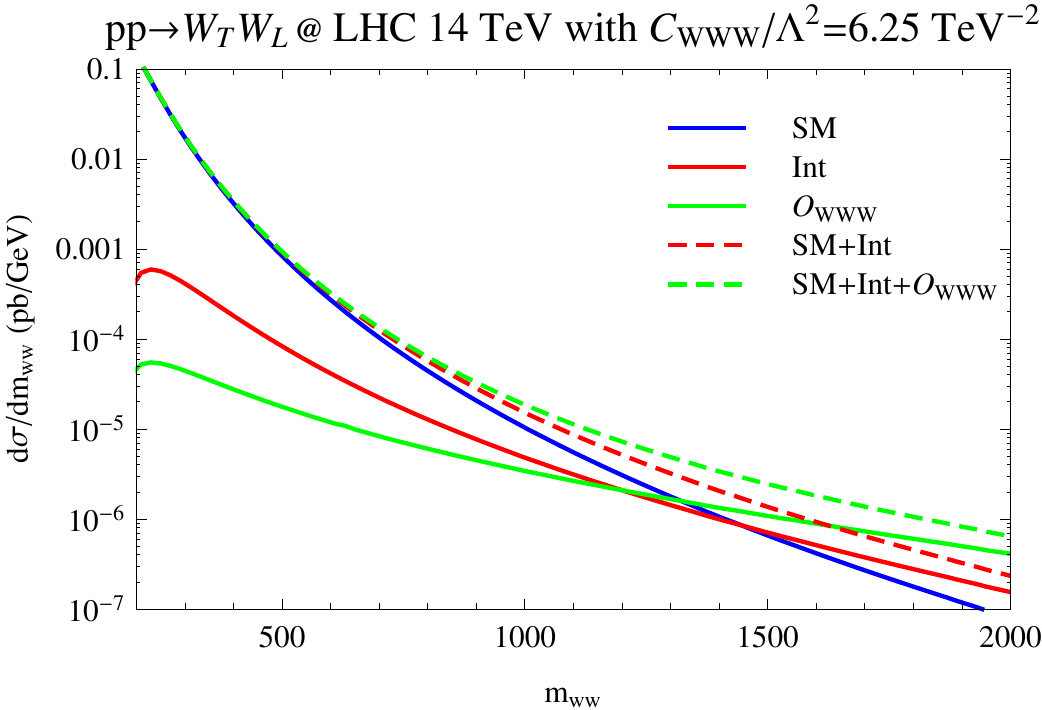}
\caption{$m_{WW}$ distributions in $W$-pair production at the 14 TeV LHC are displayed on the l.h.s. for the SM (in blue) and for the SM plus
  the dimension six operator $\mathcal{O}_{WWW}$ with $c_{WWW}/\Lambda^2=6.25$ TeV (in red). Also shown is the
  unitarity bound~\cite{Degrande:2012wf} (in
  green).  The figure on the r.h.s. shows the $m_{WW}$ distribution for the
  production of one longitudinally and one transversally polarized $W$
  boson, when considering the SM (solid blue line), only the interference between
  the SM and the dimension-six operator (solid red line), the sum
  of the two (dashed red line), only the square of the new physics
  amplitude (solid green line), and finally the total contribution from
  the SM and the dimension-six operator (dashed green line).}
\label{fig:unid6}
\end{figure}

The effective field theory is valid only below the new physics scale $\Lambda$ and no
violation of unitarity occurs in this regime.
In the regime where EFT is valid, the new physics contributions to a SM process, i.e. 
the interference of the SM amplitude with the higher-dimensional operators and the square
of the new physics amplitudes, are suppressed by increasing powers of
$1/\Lambda$,
\begin{equation}
\left| {\cal M}_{SM} + {\cal M}_{dim6}+{\cal M}_{dim8}+\ldots\right|^2 = \underbrace{\left| {\cal M}_{SM}\right|^2}_{\Lambda^0} + \underbrace{2 \Re\left({\cal M}_{SM}{\cal M}_{dim6}\right)}_{\Lambda^{-2}} +\underbrace{ \left|{\cal M}_{dim6}\right|^2 + 2 \Re\left({\cal M}_{SM}{\cal M}_{dim8}\right)}_{\Lambda^{-4}} +\ldots 
\end{equation}
For illustration we show in Fig.~\ref{fig:unid6} the invariant mass distribution of the $W$-pair, $m_{WW}$, 
produced at the 14 TeV LHC, with and without the contribution of 
the dimension six operator ${\cal O}_{WWW}$ of Eq.~\ref{opTGC}. As can be seen on the l.h.s., the
prediction for $m_{WW}$ including ${\cal O}_{WWW}$ is well below the unitarity bound   
~\cite{Degrande:2012wf} for this process in the relevant energy regime.
However, as illustrated on the r.h.s.,  the contributions of this operator to the amplitude squared
for $W_LW_T$ production reach similar magnitude at
$m_{WW}\approx1.3$ TeV and above this energy the $1/\Lambda^4$ suppressed term overtakes the $1/\Lambda^2$ suppressed contribution. Clearly, the $1/\Lambda$ expansion is only valid below this energy.

For dimension eight operators, the effect from unitarity violation
typically sets in earlier due to the higher exponent in $\Lambda$ in the
denominator. Hence, the task to avoid unphysical contributions from
regions where unitarity is violated becomes more important. In these
regions the EFT expansion in terms of suppressed additional
contributions to the SM part, our starting point, is no longer valid, as
each order becomes similarly important.

In experimental searches one has to ensure that the sensitivity on
anomalous gauge couplings is not driven by parameter regions where
unitarity is violated. As nature will ensure unitarity conservation in
the full model, such results would not be meaningful. Thereby, one can
take advantage of the fact that only energies up to the center-of-mass
energy of the collider are probed. For hadron colliders like the LHC,
the steep fall-off of the parton distribution functions means that the
effective probed energy range is even smaller, as the expected number of
signal events will be smaller than one above a certain energy and
therefore this region will not contribute.
However, if the bound for unitarity violation is lower than that, some
method to ensure that no sensitivity comes from this energy range needs
to be employed. One possibility is to use appropriate experimental cuts.
However, often processes will contain neutrinos and so the full
reconstruction of the partonic energy is not possible. Another option
are form factors. These are introduced to model an energy-dependent
cutoff, which in the full theory would be accomplished by new-physics
states at the scale $\Lambda$, which have been integrated out in the EFT
description. Various options are possible, for example a sharp cut-off
of the higher-dimensional contributions at a fixed energy scale, or a
dipole-like form factor as used in \VBFNLO{}, that gives a smoother
cut-off. The exact choice depends on the full model, so for an effective
theory description all choices are equally well motivated from the
theory side. The last possibility to ensure no unitarity violation
happens is a unitarity projection, like the $K$-matrix method
implemented in WHIZARD. There the amplitude $A$ is moved onto the
unitarity circle along a line connecting $\Re(A)$ and the imaginary unit
$i$. Physically, this corresponds to introducing an infinitely heavy and
wide resonance.  This scheme maximizes the contributions from anomalous
couplings while ensuring unitarity for all energies.

\clearpage

\section{Predictions for multi-boson processes with non-standard couplings}
\label{sec:pred}

\subsection{\textsc{ MadGraph5}}

\textsc{MadGraph5} \cite{Alwall:2011uj} is a suite of programs related
to the numerical evaluation of the matrix element
\cite{Hirschi:2011pa,Artoisenet:2010cn,Frederix:2009yq}.  In
particular, the tool is able to compute the cross section and to
generate events at leading order accuracy \cite{Maltoni:2002qb}.  It
also contains an interface to \textsc{Pythia6} \cite{Sjostrand:2006za}
to generate inclusive samples at LO+PS accuracy via the CKKW
matching/merging scheme \cite{Catani:2001cc,Alwall:2007fs}.
Additionally, a public beta version (2.0.0beta4) of the code allows to
perform the computation at next-to-leading order accuracy in QCD
matched to a parton shower via the \textsc{aMC@NLO} module
\cite{Alwall:2013}.  As in the LO mode, there is no predefined list of
processes, \textsc{aMC@NLO} is able to generate fully automatically an
optimized way to evaluate the matrix element and the associate
phase space integration.

At leading order, the program has been designed to be fully model
independent. In addition to its own model format, \textsc{MadGraph5}
contains an interface to support a model written in the \textsc{UFO}
convention \cite{Degrande:2011ua}. This model is by essence fully
generic and not tied to any Monte Carlo generator. Unfortunately,
MadGraph has some small limitation on the model that can be imported
via this format.  First, \textsc{MadGraph5} does not support spin
larger than 2. Secondly, the color module does not support representation 10
or higher, but includes the sextet and the support for the fully
anti-symmetric color-structure. On the other hand, there are no
limitations on the Lorentz structure allowed for a given interactions
and in particular on the number of particles. Indeed
\textsc{MadGraph5} calls the \textsc{ALOHA} package
\cite{deAquino:2011ub} in order to create the helicity amplitude
routine \cite{HELAS} that are needed for the efficient evaluation of
the matrix element.  As a small exception, \textsc{MadGraph5} does not
support multi-fermion interactions in presence of fermion-flow
violation, but all other type of multi-fermion interactions are
supported including the case of identical fermions.  A recent extension
of the \textsc{UFO} and \textsc{ALOHA} package
\cite{Christensen:2013aua} allows \textsc{MadGraph5} to support
user-defined propagators as well as form factors.

Writing a \textsc{UFO} model is obviously somewhat tedious,
fortunately various packages allow to create models automatically
for a large class of local theories.  This format is currently
supported by \textsc{FeynRules}
\cite{Christensen:2008py,Christensen:2009jx} and \textsc{SARAH}
\cite{Staub:2012pb}, and is planned to be supported by \textsc{LanHep}
\cite{Semenov:2008jy} as well.  An extension of the \textsc{UFO} model
for next-to-leading accuracy is on its way as well as a \textsc{FeynRules}
interface to create models automatically \cite{Celine_r2}.
 
One key feature of \textsc{MadGraph5} at leading order is that one can
easily specify the decay chain structure associated to a
production process. In such cases, \textsc{MadGraph5} is able to generate
events with up to 16 particles in the final state including full
spin correlations and off-shell effects\footnote{Since such
  computations are stricto-senso only valid in the narrow-width
  approximation, a customizable cut is added to forbid the decaying
  particles to be too far off-shell.}. An alternative, which is especially
useful at NLO, consists of generating the production process at parton-level
without decay, and then using the \textsc{MadSpin} package
\cite{Artoisenet:2012st} to generate the decay also with full
spin-correlations and off-shell effects.

\textsc{MadGraph5} contains also a large class of options concerning
the parton-level cuts and beam parametrizations.  For example, it
supports polarized beams and beamstrahlung, \textsc{LHAPDF}.  Finally, the
\textsc{MadGraph5} interface is designed to be user friendly and
contains a built-in tutorial to facilitate the apprentissage procedure
and can install (and link) fully automatically a series of external
codes (e.g., \textsc{Pythia6} \cite{Sjostrand:2006za} ,
\textsc{MadAnalysis} \cite{Conte:2012fm}, \textsc{Delphes}
\cite{deFavereau:2013fsa}).  With all those features,
\textsc{MadGraph5} is a very flexible tool which can describe
efficiently and precisely a large class of phenomenological processes,
and in the electroweak sector in particular.  It is therefore often a tool of
choice for the study of the physics potential of future accelerators.

\subsection{\VBFNLO}
\label{sec:vbfnlo}

\VBFNLO~\cite{Arnold:2012xn,Arnold:2011wj,Arnold:2008rz} is a flexible
parton-level Monte-Carlo generator for processes with electroweak
bosons. It allows the simulation of vector-boson fusion processes with the
production of a Higgs boson or one or two massive gauge bosons as well
as the production of two or three electroweak gauge bosons, including
final states with photons. All these processes are implemented at
next-to-leading order in the strong coupling constant. Furthermore,
gluon-fusion production of Higgs plus two jets and of two electroweak
bosons is implemented at the leading one-loop level. The program allows
to place arbitrary cuts on the final-state particles and implements
various scale choices. Any available PDF set can be used via a link to
LHAPDF\cite{Whalley:2005nh}. Events can be written out both in weighted
and unweighted form and the LHE~\cite{Alwall:2006yp} as well as the
HepMC~\cite{Dobbs:2001ck} format.

All processes include fully leptonic decays of the gauge bosons, where
the user can choose whether a particular final state is desired or all
combinations, with or without the third generation, should be
summed over and included in the event file.
Off-shell effects including contributions from virtual photons instead
of $Z$ bosons are taken into account, while Pauli-interference effects
for identical charged leptons are neglected.  Furthermore, for the
production of two massive gauge bosons, both direct and via vector-boson
fusion, and the triboson process $W^+W^-Z$, semi-leptonic decays are
available as well, where one boson decays into a quark pair,
while the other ones still decay leptonically.
Again either a specific flavor final state, only first- and
second-generation quarks, or all light quarks including bottom quarks
can be chosen. An extension to the other triboson processes is planned
for the future. 

Anomalous triple and quartic gauge couplings are implemented for all
vector-boson-fusion processes with production of one or two gauge
bosons, and all diboson and triboson
processes~\cite{FeiglDipl,SchlimpertDipl}. The operator structure has
already been described in Section~\ref{sec:param}. Note that the
considered operators do not give any contributions to the diboson
processes with two neutral particles in the final state, i.e.\ $ZZ$,
$Z\gamma$ and $\gamma\gamma$. In all processes a form factor
\begin{equation}
F = \left(1+\frac{s}{\Lambda}\right)^{-p}
\end{equation}
can be applied to ensure unitarity at high energies~\cite{Barger:1990py,
Baur:1987mt, Baur:1988qt, Gounaris:1993fh, Gounaris:1994cm}. Here $s$ is
a universal scale for each phase-space point, taken to be equal to the
squared invariant mass of the produced bosons and $\Lambda$ and $p$ are
free parameters describing the mass scale of the cut-off and the power
of the damping, respectively. $p$ should be chosen to be at least $1$
for the dimension-6 and $2$ for the dimension-8 operators to possess the
required damping at high energies.

Additionally, a dedicated form factor tool can be downloaded from the
\VBFNLO{} web site~\cite{VBFNLO_formfac}. The tool calculates
on-shell $VV$ scattering and computes the lowest ($J=0$) contribution to
the partial wave decomposition of the amplitude. The unitarity criterion
is that the real part of this contribution must be below 0.5~\cite{Barger:1990py}.
We check each possible combination in $VV \rightarrow VV$ where $V=W$,
$Z$, $\gamma$ separately as well as the combination of all channels with
the same electrical charge of the $VV$ system. After reading the
anomalous coupling parameters from an input file, the output of the program
then consists of the partonic center-of-mass energy for each channel
where unitarity is first violated. This is performed both for the 
helicity combination giving the largest constraint and the most
restrictive linear combination obtained by diagonalizing the $T$-matrix. 
Additionally, a value for $\Lambda$ is calculated for each case that
just ensures tree-level unitarity up to the given energy, taking the
exponent $p$ and the maximum considered energy set in the input file.

\subsection{The event generator WHIZARD}
\label{sec:whizard}

WHIZARD~\cite{Kilian:2007gr} is a Monte Carlo event generator for
hadron and lepton colliders. The most recent public version is
2.1.1, while an $\alpha$ release of the new version 2.2 will come
out later this summer. WHIZARD contains the optimizing matrix 
element generator O'Mega~\cite{Moretti:2001zz}. O'Mega has been
written in the functional language OCaml that allows for great
versatility and flexibility. It uses the concept of directed acyclical
graphs (DAGs) to generate amplitudes that are optimal in the sense
that all redundancies due to common subamplitudes and gauge invariance
have been avoided. On top of that, a common subexpression elimination
for equivalent flavor combinations is foreseen~\cite{O2}. QCD quantum
numbers are treated using the color flow
formalism~\cite{Kilian:2012pz}, that is ideally suited for
transferring the color information to the parton shower
(WHIZARD has its own $k_T$-ordered and analytic parton
showers~\cite{Kilian:2011ka}, but no hadronization). O'Mega
supports all spins from scalars up to spin 3/2~\cite{Reuter:2002gn}
and tensor particles. A large library of vertex functions is
supported for 3- and 4-point interactions as well as for
higher-dimensional operators. Completely general Lorentz structures
will be supported in the upcoming version, 2.2~\cite{O2}. WHIZARD is
particularly specialized an beyond the SM (BSM) models, containing a
large number of implemented models ranging from a variety of SUSY
models, Little Higgs models, extra-dimensional models to models with
anomalous couplings. An interface to the Lagrangian to Feynman rules
converter FeynRules allows for the inclusion of basically arbitrary
QFT-based BSM models~\cite{Christensen:2010wz}. Both
O'Mega and WHIZARD have a large intrinsic testsuite that guarantees
the inner consistency and prevents regressions during the development,
e.g. there are Ward- and Slavnov-Taylor identities being
checked~\cite{Ohl:2002jp}. 

Phase-space integration is performed by an adaptive multi-channel
Monte-Carlo integration provided by the subpackage
VAMP~\cite{Ohl:1998jn}. The new version of WHIZARD also contains
alternative integration methods that are e.g. better suited for simple
decay processes. The WHIZARD core that has been recasted in a very
modern, modularized and object-oriented form in Fortran2003 steers the
matrix element generation, compilation, phase space generation and
interfacing to external libraries for PDFs, event formats, and
hadronization. The input to WHIZARD happens through a universal
scripting language SINDARIN, which is a very self-contained
user-friendly syntax as input method to define processes, scales,
cuts, and analyses. This input syntax allows to define arbitrary
kinematical expressions for cuts and scales. 

One of the main fields of WHIZARD applications (as in the context here)
is for electroweak physics, particularly anomalous couplings and new
resonances in the EW
sector~\cite{Beyer:2006hx,Alboteanu:2008my,Kilian:prep}. Other areas of
applications not relevant in the context of the EW Snowmass White Paper
(QCD, other BSM, ILC physics, etc.) are left out here for brevity.

\subsection{The Role of Higher-order Corrections}
\label{sec:ho}

Higher-order corrections play an important role for accurate predictions
at the LHC. In this section we study the impact of NLO QCD corrections
in vector-boson fusion and triboson processes and how they impact the
extraction of anomalous quartic gauge couplings. As example of these two
process classes we take the processes $W^+W^+jj$ and $W^+\gamma\gamma$,
respectively. The NLO results including anomalous QGCs presented in Sections~\ref{sec:wwjjvbfnlo} and ~\ref{sec:wggvbfnlo} have been obtained with \VBFNLO{}.
We discuss the impact of a parton shower on the example of $W^+W^+jj$ production with {\tt POWHEG+PYTHIA}~\cite{Jager:2011ms} in Section~\ref{sec:wwjjpowheg}.
Finally, in Section~\ref{sec:tribosonew} we discuss the impact of NLO
electroweak corrections in triboson processes.

\subsubsection{Vector-boson-fusion process $W^+W^+jj$ with VBFNLO}
\label{sec:wwjjvbfnlo}

The production of a vector-boson pair via vector-boson
fusion~\cite{Jager:2006zc, Jager:2006cp, Bozzi:2007ur, Jager:2009xx,
Denner:2012dz} has a characteristic signature of two high-energetic,
so-called tagging jets in the forward region of the detector, which are
defined as the two jets with the largest transverse momentum. This can
be exploited experimentally by requiring that there is a large rapidity
separation ($\Delta\eta_{jj}>4$) between the tagging jets, they are in
opposite detector hemispheres ($\eta_{j_1} \times \eta_{j_2} < 0$) and
they possess a large invariant mass ($M_{jj} > 600$ GeV). Additional
central jet radiation at higher orders is strongly suppressed due to the
exchange of a color-singlet in the t-channel, in contrast to typical
QCD-induced backgrounds. Higher-order corrections are typically small,
below the 10\% level, and reduce the residual scale uncertainty to about
2.5\%. Choosing the momentum transfer between an incoming and an
outgoing parton along a fermion line proves to be particularly
advantageous, as then also corrections to important distributions are
small and flat over the whole range. 

\begin{figure}
\begin{center}
\includegraphics[height=0.3\textwidth]{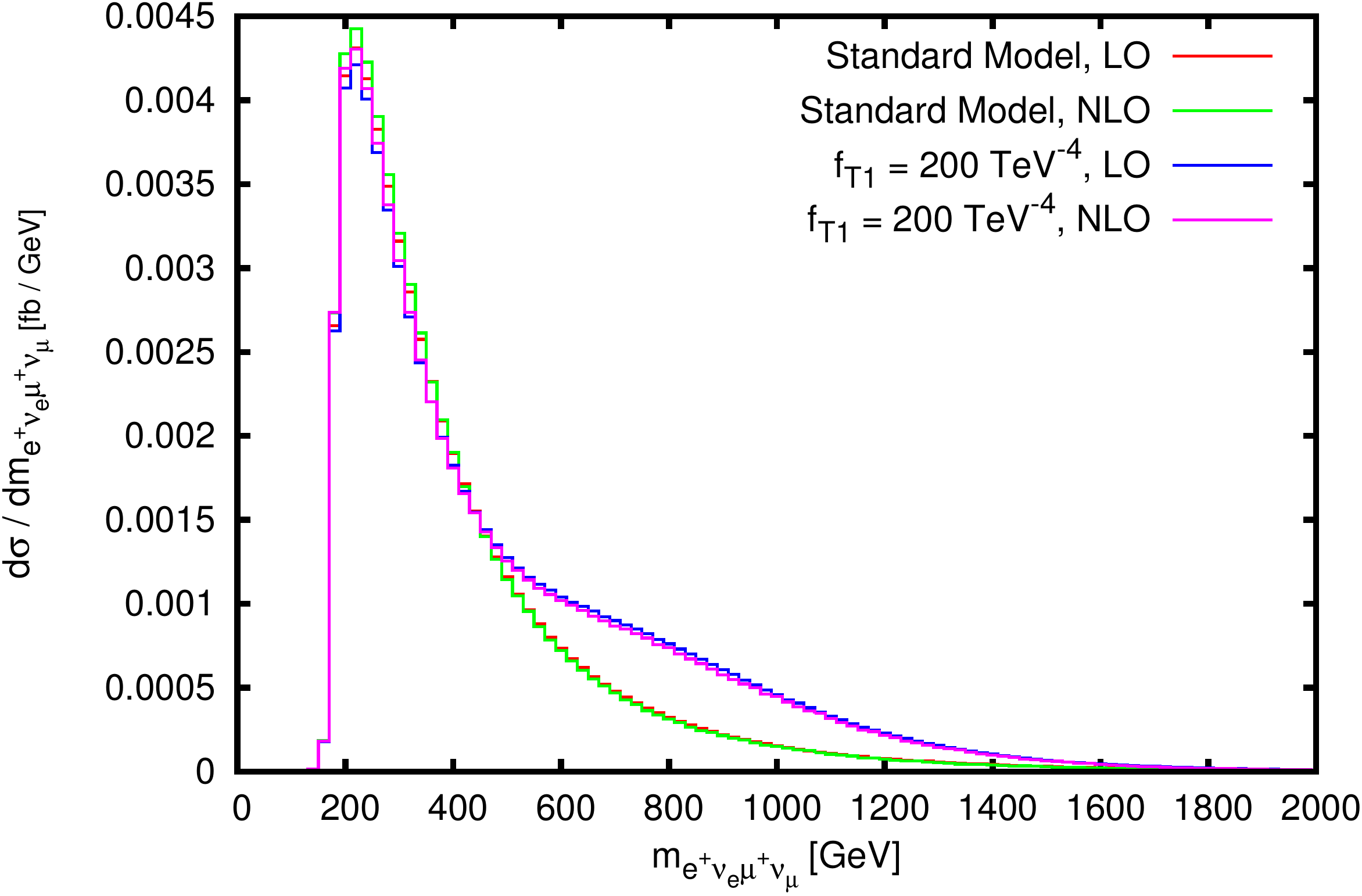} \quad
\includegraphics[height=0.3\textwidth]{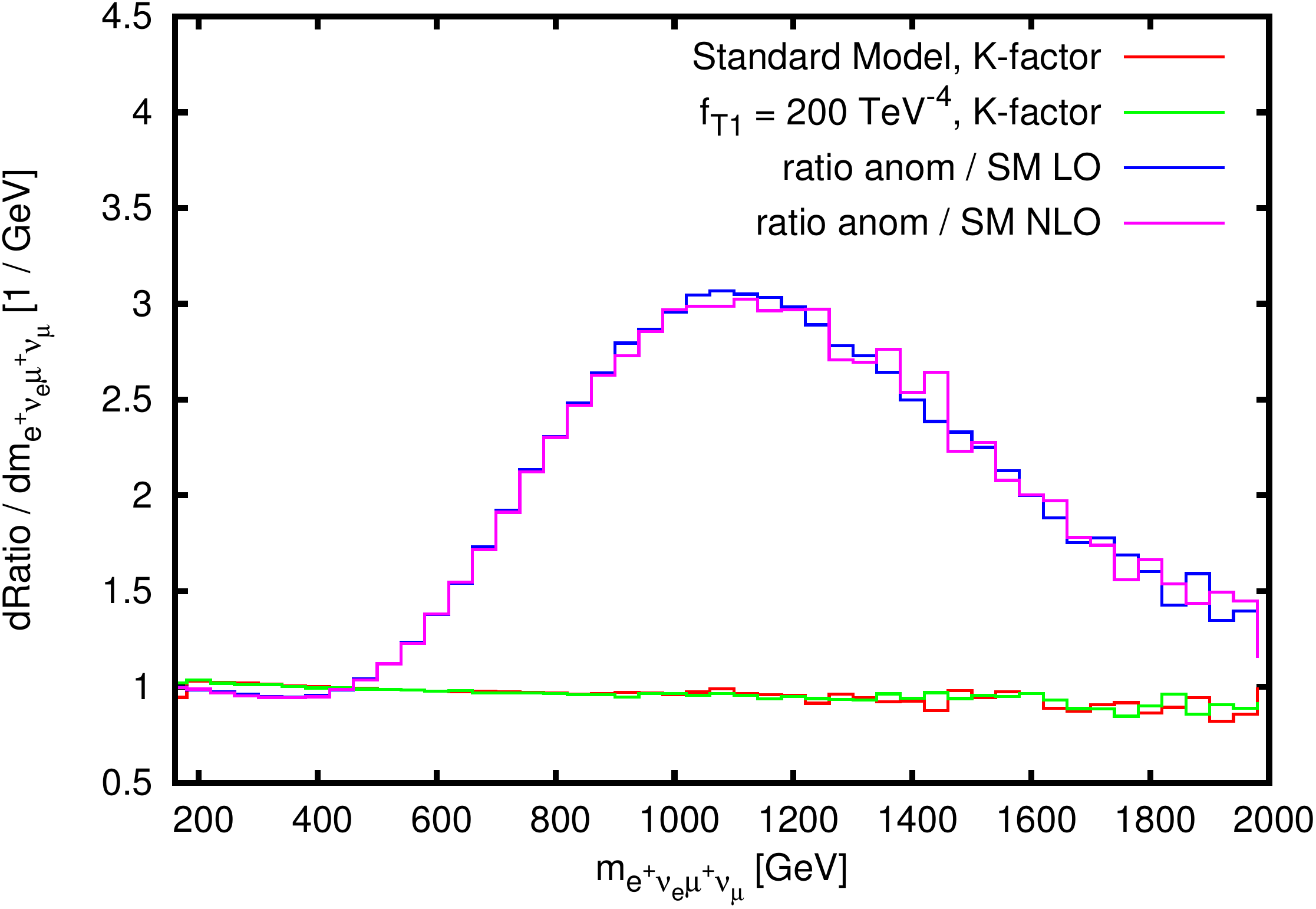}
\end{center}
\caption{Invariant-mass distribution of the two lepton, two neutrino
system. \textit{Left:} Differential cross section for the SM and with
anomalous coupling $T_1$ at LO and NLO. \textit{Right:} Differential
K-factors for the SM and with anomalous coupling as well as the
cross-section ratio between anomalous coupling and SM for LO and NLO.}
\label{NLO:VBF:mVV}
\end{figure}

As example we take the process $pp \rightarrow e^+ \nu_e \mu^+ \nu_\mu
jj$ with anomalous coupling $\frac{f_{T,1}}{\Lambda^4} = 200
\text{ TeV}^{-4}$ and formfactor scale $\Lambda=1188$ GeV and exponent
$p=4$. The results for the total cross sections at LO and NLO are shown
in Tab.~\ref{NLO:VBF:cs}.
\begin{table}
\begin{tabular}{l|cc}
& $\sigma_{\text{LO}}$ & $\sigma_{\text{NLO}}$ \\\hline
SM          & 1.169 fb & 1.176 fb \\
anom.coupl. & 1.399 fb & 1.388 fb \\
\end{tabular}
\caption{Total cross sections at LO and NLO for the process $pp
\rightarrow e^+ \nu_e \mu^+ \nu_\mu jj$ in the SM and with anomalous
coupling $\frac{f_{T,1}}{\Lambda^4} = 200 \text{ TeV}^{-4}$. Statistical
errors from Monte Carlo integration are below the per mille level.}
\label{NLO:VBF:cs}
\end{table}
Switching on the anomalous couplings increases the cross section by just
under 20\%, and NLO QCD corrections hardly change this number. This can
also be seen in Fig.~\ref{NLO:VBF:mVV} where we show the differential
distribution with respect to the invariant mass of the two leptons and
the two neutrinos.
In the left-hand plot we present the differential cross section in the
SM and with anomalous coupling switched on both at LO and NLO. Similar
to the integrated cross section, the difference between LO and NLO is
small in both cases. In contrast the anomalous couplings yield a
positive contribution to the cross section over the SM, which starts at
an invariant mass of about 500 GeV, before the formfactor, introduced to
preserve unitarity, damps the contributions again at higher invariant
masses. On the right-hand side we present two groups of ratios. The
differential K factor is flat and close to one both for the SM and the
anomalous coupling scenario. The second set shows the ratio of
differential anomalous-coupling over SM cross section both at LO and
NLO. The two curves agree well and show enhancements of the cross
section up to a factor of three. Hence, in this process higher-order
corrections do not influence the extraction of anomalous couplings.

\subsubsection{Vector-boson-fusion process $W^+W^+jj$ in the \POWHEGBOX}
\label{sec:wwjjpowheg}

NLO-QCD calculations are a crucial prerequisite for precision analyses
at the LHC, reducing theoretical uncertainties associated with hard
scattering processes significantly. On the other hand, a realistic
description of the additional hadronic activity that occurs in any
collider environment crucially relies on parton-shower Monte Carlo
generators such as \HERWIG{}~\cite{Corcella:2000bw} or
\PYTHIA{}~\cite{Sjostrand:2006za}. The perturbative accuracy of these
programs is, however, limited to leading logarithmic accuracy.
The most realistic yet accurate predictions available to date for
processes with many particles in the final state are thus obtained by
combining NLO-QCD calculations for the hard scattering with parton
shower programs, for example in the framework of the \POWHEG{}
formalism~\cite{Nason:2004rx,Frixione:2007vw}.  Such a matching can be
performed with the help of the \POWHEGBOX{}~\cite{Alioli:2010xd}, a
repository that provides all process-independent building blocks of
the matching procedure, while process-specific elements have to be
provided by the user.

Building on existing NLO-QCD calculations
\cite{Figy:2003nv,Oleari:2003tc,Jager:2009xx,Jager:2006zc}, recently
various VBF processes have been implemented in the
\POWHEGBOX{}~\cite{Nason:2009ai,Jager:2012xk,Schissler:2013nga,Jager:2011ms,Jager:2013mu}. The
code developed is publicly available from the project webpage, {\tt
  http://powhegbox.mib.infn.it/}, and can be tailored to the user's
needs for any dedicated study.
In order to assess the impact of parton-shower effects on NLO-QCD
predictions for VBF-induced $W^+W^+jj$ production at the LHC,
numerical analyses for a representative setup have been performed for
the $e^+\nu_e\mu^+\nu_\mu jj$ final state~\cite{Jager:2011ms}.  At a
collision energy of $\sqrt{s}=7$~TeV, the MSTW2008 parton distribution
functions~\cite{Martin:2009iq} are used for incoming protons and the
{\tt FASTJET} package~\cite{Cacciari:2005hq} for the reconstruction of
jets via the $k_T$~algorithm with a resolution parameter of
$R=0.4$. Events are showered with {\PYTHIA~6.4.21}, including
hadronization corrections and underlying event with the Perugia~0
tune.
At least two hard jets are required with $p_{T,j}\geq 20$~GeV and
$|y_j|\leq 4.5$, well-separated from each other such that
$|y_{j_1}-y_{j_2}|>4$, $y_{j_1}\times y_{j_2}<0$, and
$M_{j_1j_2}>600$~GeV.  In addition, an $e^+$ and a $\mu^+$ with
$p_{T,\ell}\ge 20$~GeV, $|y_\ell|\le 2.5$, $\Delta R_{j\ell}\ge
0.4$, $\Delta R_{\ell\ell}\ge 0.1$, located between the two tagging
jets, are requested. For the renormalization and factorization scales
dynamical choices bound to the kinematics of the underlying Born
configuration are made.

\begin{figure}
\begin{center}
\includegraphics[height=0.3\textwidth]{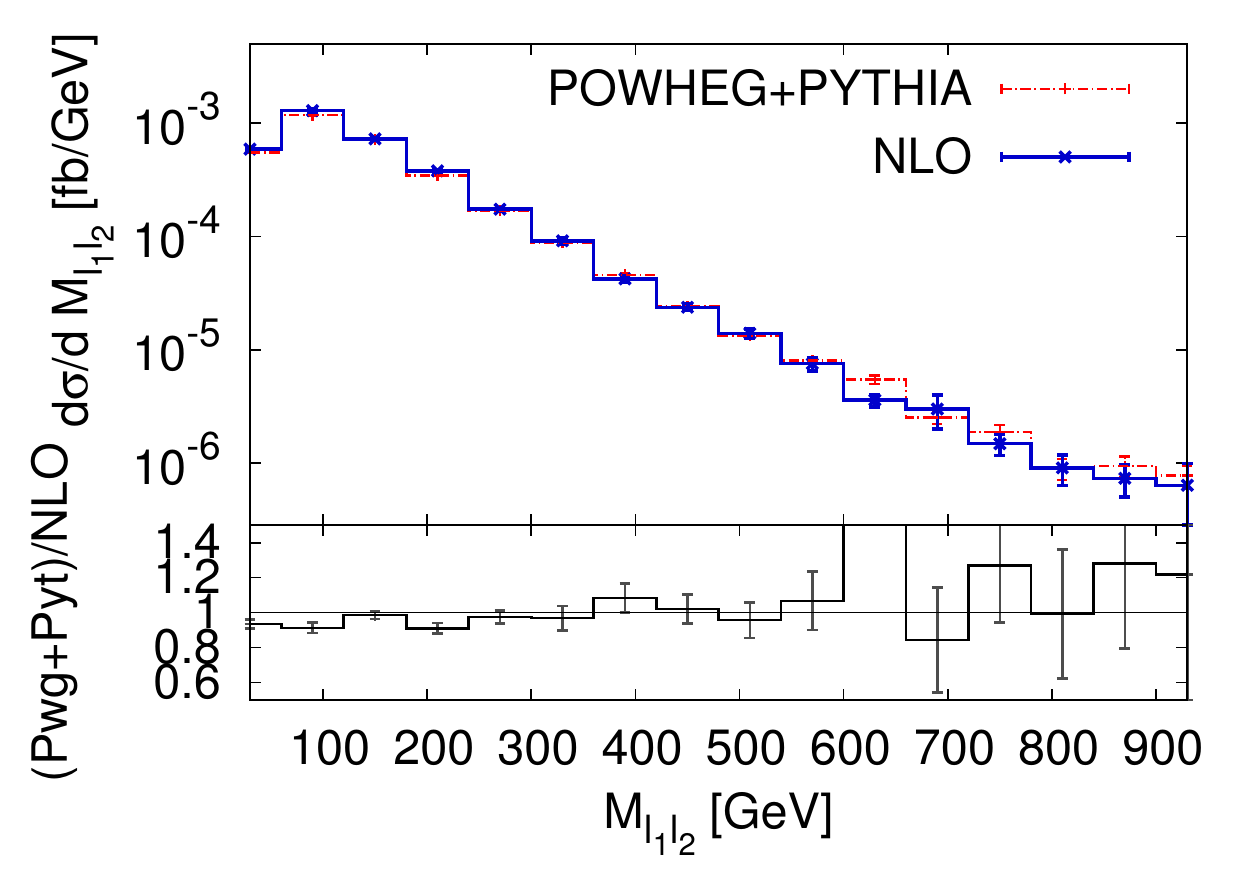} \quad
\includegraphics[height=0.3\textwidth]{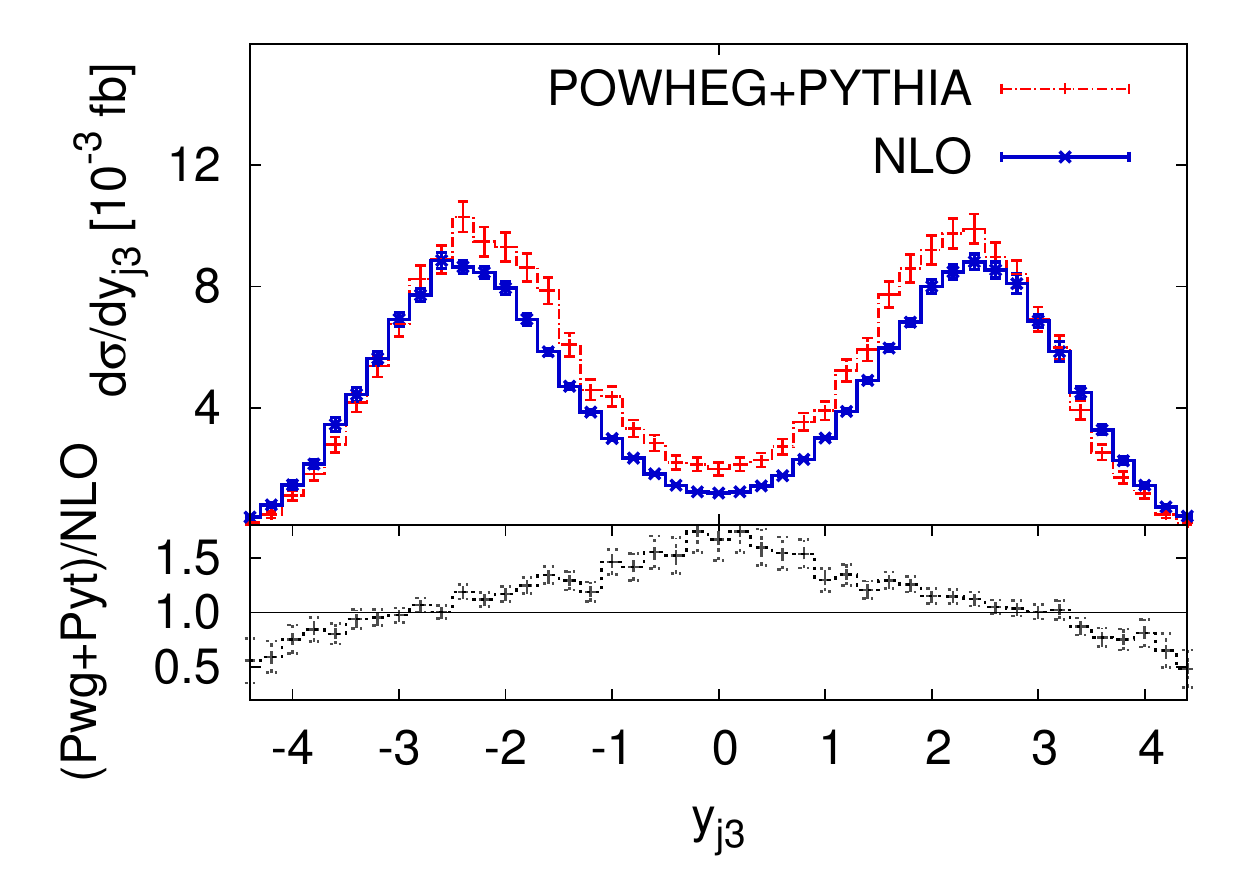}
\label{fig:vbf-wpp-powheg}
\caption{Invariant mass distribution of the charged lepton pair (left)
  and rapidity distribution of the third jet (right) in VBF-induced
  $e^+\nu_e\mu^+\nu_\mu jj$ production at the LHC with
  $\sqrt{s}=7$~TeV and the selection cuts described in the text.  The
  lower panels show the respective ratios of the {\tt POWHEG+PYTHIA}
  and the NLO-QCD results. Horizontal bars indicate statistical errors
  in each case. }
\end{center}
\end{figure}

In this setup distributions related to the tagging jets or the hard
leptons turn out to be rather insensitive to parton-shower effects. As
illustrated by Fig.~\ref{fig:vbf-wpp-powheg}~(left panel) for the
invariant mass distribution of the charged-lepton pair, the NLO-QCD
and the {\tt POWHEG+PYTHIA} results are very similar, both in
normalization and shape.
More pronounced effects of the parton shower occur in observables
related to the emission of an extra hard jet,
c.f.~Fig.~\ref{fig:vbf-wpp-powheg}~(right panel) for
$d\sigma/dy_{j_3}$. When the rapidity distribution of a third jet is
used in order to estimate central-jet veto efficiencies this effect
should be carefully taken into account.

\subsubsection{Triboson process $W^+\gamma\gamma$ with VBFNLO}
\label{sec:wggvbfnlo}

The second group of process where anomalous quartic gauge couplings can
be tested are the triboson processes~\cite{Lazopoulos:2007ix,
Hankele:2007sb, Campanario:2008yg, Binoth:2008kt, Bozzi:2009ig,
Bozzi:2010sj, Baur:2010zf, Bozzi:2011wwa, Bozzi:2011en, Bozzi:2012mh,
Campbell:2012ft}. The quartic vertex enters via an $s$-channel vector
boson, which decays into three vector bosons, while diagrams with two or
three bosons attached to the quark line as well as non-resonant
contributions form an irreducible background. These processes have been
shown to possess quite large K factors, typically between 1.5 and 1.8,
mostly due to the additional quark-gluon--induced production processes
first entering in the real-emission process. They also have a
considerable scale dependence.  While the dependence on the
factorization scale can be reduced by NLO QCD corrections, the strong
coupling constant first enters in the real emission part and therefore
shows a large variation with the scale.  

The example process we are considering here is $pp \rightarrow e^+ \nu_e
\gamma \gamma$~\cite{Baur:2010zf, Bozzi:2011wwa}. In this process the K
factor with a numerical value of about 3 is particularly large. This is
due to the fact that the SM amplitude vanishes when the two photons are
collinear and $\cos{\theta_W} = \frac13$, where $\theta_W$ is the angle
between the $W$ and the incoming quark in the partonic center-of-mass
frame.  This so-called radiation zero~\cite{Brown:1982xx, Baur:1993ir,
Baur:1997bn} is spoiled by the extra jet emission at NLO, therefore
giving huge K factors in these phase-space regions. The numerical values
for the integrated cross section are tabulated in
Table~\ref{NLO:VVV:cs}.
\begin{table}
\begin{tabular}{l|cc}
& $\sigma_{\text{LO}}$ & $\sigma_{\text{NLO}}$ \\\hline
SM          & 1.124 fb & 3.674 fb \\
anom.coupl. & 1.216 fb & 3.787 fb \\
\end{tabular}
\caption{Total cross sections at LO and NLO for the process $pp
\rightarrow e^+ \nu_e \gamma \gamma$ in the SM and with anomalous
coupling $\frac{f_{T,6}}{\Lambda^4} = 2000 \text{ TeV}^{-4}$.
Statistical errors from Monte Carlo integration are below the per mille
level.}
\label{NLO:VVV:cs}
\end{table}
As anomalous coupling we choose the operator $T_6$ with
$\frac{f_{T,6}}{\Lambda^4} = 2000 \text{ TeV}^{-4}$, formfactor scale
$\Lambda=1606$ GeV and exponent $p=4$. 

\begin{figure}
\begin{center}
\includegraphics[height=0.3\textwidth]{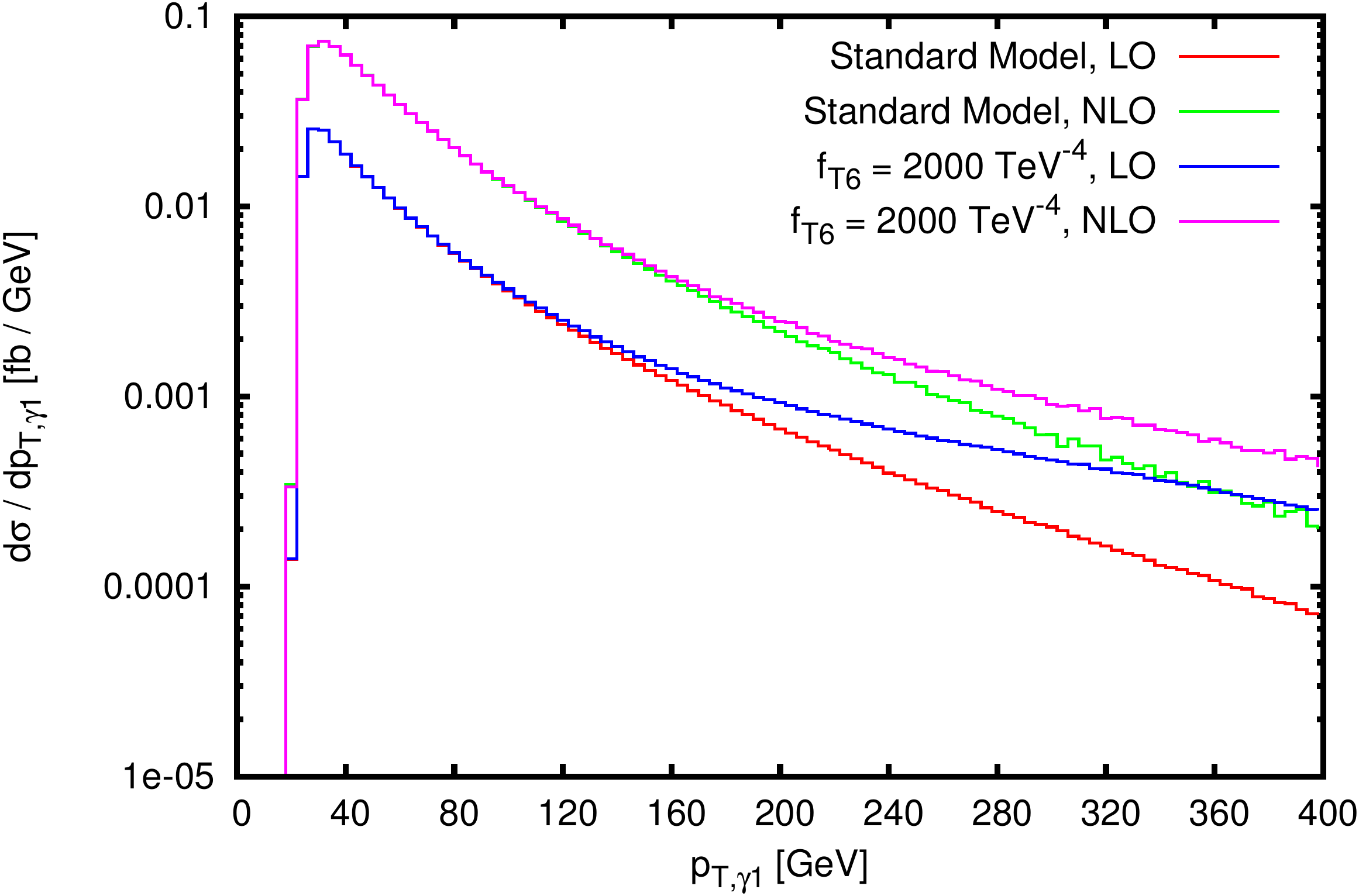} \quad
\includegraphics[height=0.3\textwidth]{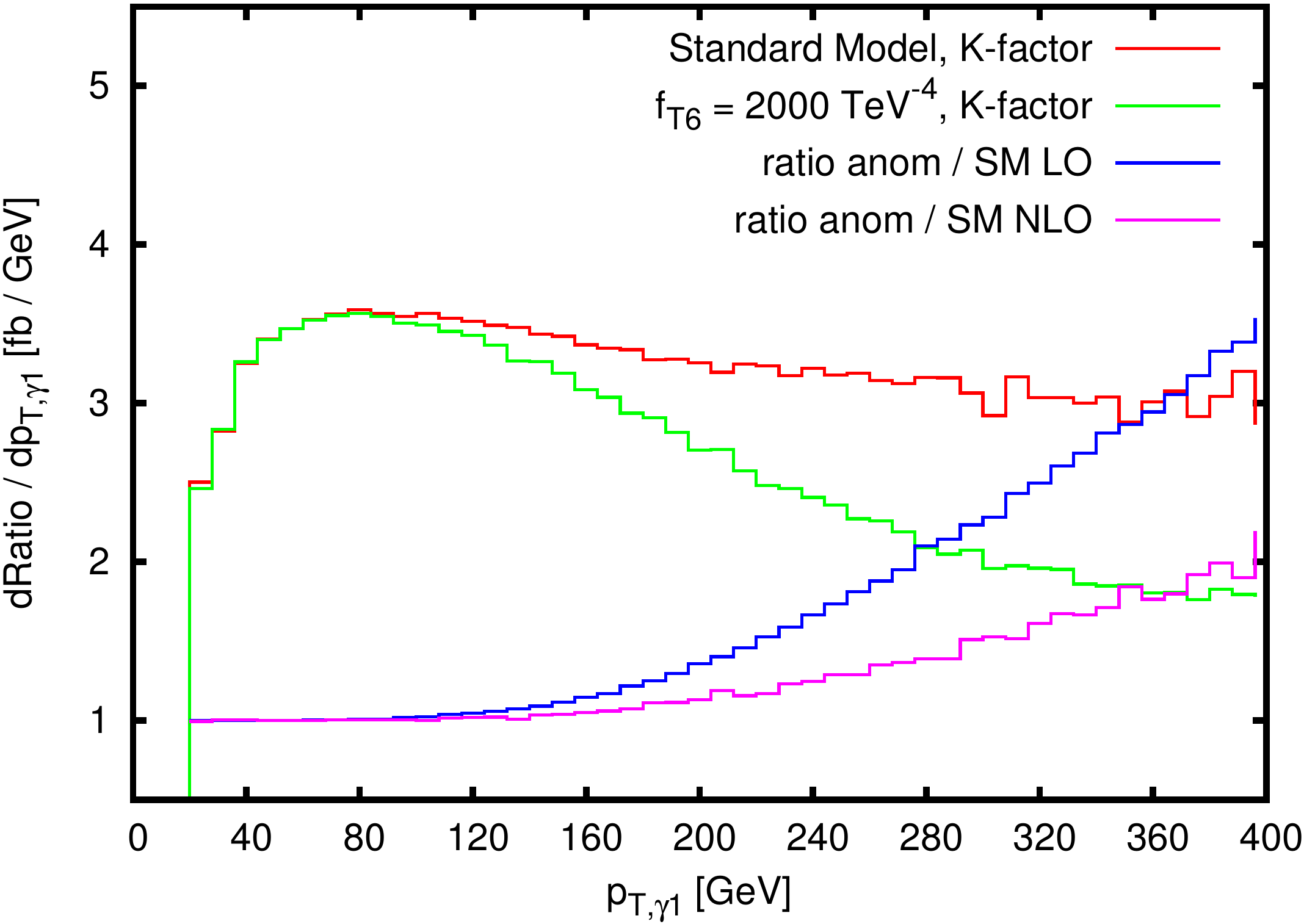}
\end{center}
\caption{Transverse-momentum distribution of the harder photon.
\textit{Left:} Differential cross section for the SM and with anomalous
coupling $T_6$ at LO and NLO. \textit{Right:} Differential K-factors for
the SM and with anomalous coupling as well as the cross-section ratio
between anomalous coupling and SM for LO and NLO.} 
\label{NLO:VVV:pTA1}
\end{figure}

Turning to differential distributions, we show the transverse momentum
distribution of the harder photon in Figure~\ref{NLO:VVV:pTA1}.
The left-hand side shows again the differential integrated cross
section. Both the SM and the anomalous-coupling scenario show
differential NLO cross sections which are significantly larger than
their LO counterpart. Contributions from anomalous couplings start to
contribute for transverse photon momenta above 100 GeV and their
relative size becomes gradually larger when going to higher momenta as
expected.

On the right-hand side one can see that the K-factor behavior differs
for the SM and the anomalous coupling scenario. While, in the SM, the K
factor is almost constant and only slightly decreases when going to
larger transverse momenta, there is a much stronger decrease when 
anomalous couplings are switched on. At the high end of the shown range,
the K factor has reached a value of around 1.8, which is the number
typically observed in other triboson processes involving $W$s. As the
effect of the anomalous coupling increases, the cancellation between
different amplitudes gets gradually destroyed and the radiation zero
filled up. Only the effects from additional jet radiation remain,
yielding the smaller K factor. 

\begin{figure}
\begin{center}
\includegraphics[height=0.3\textwidth]{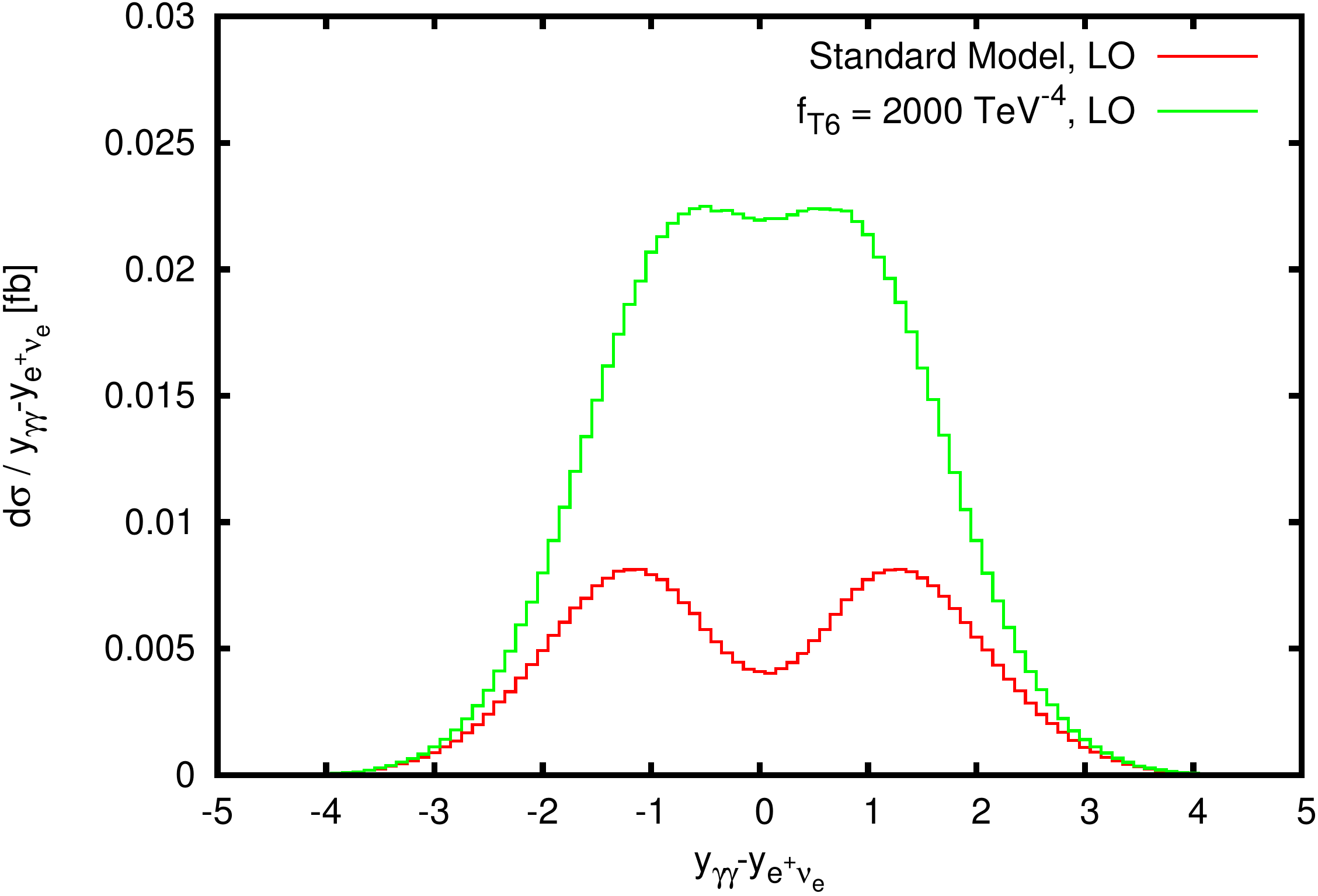} \quad
\includegraphics[height=0.3\textwidth]{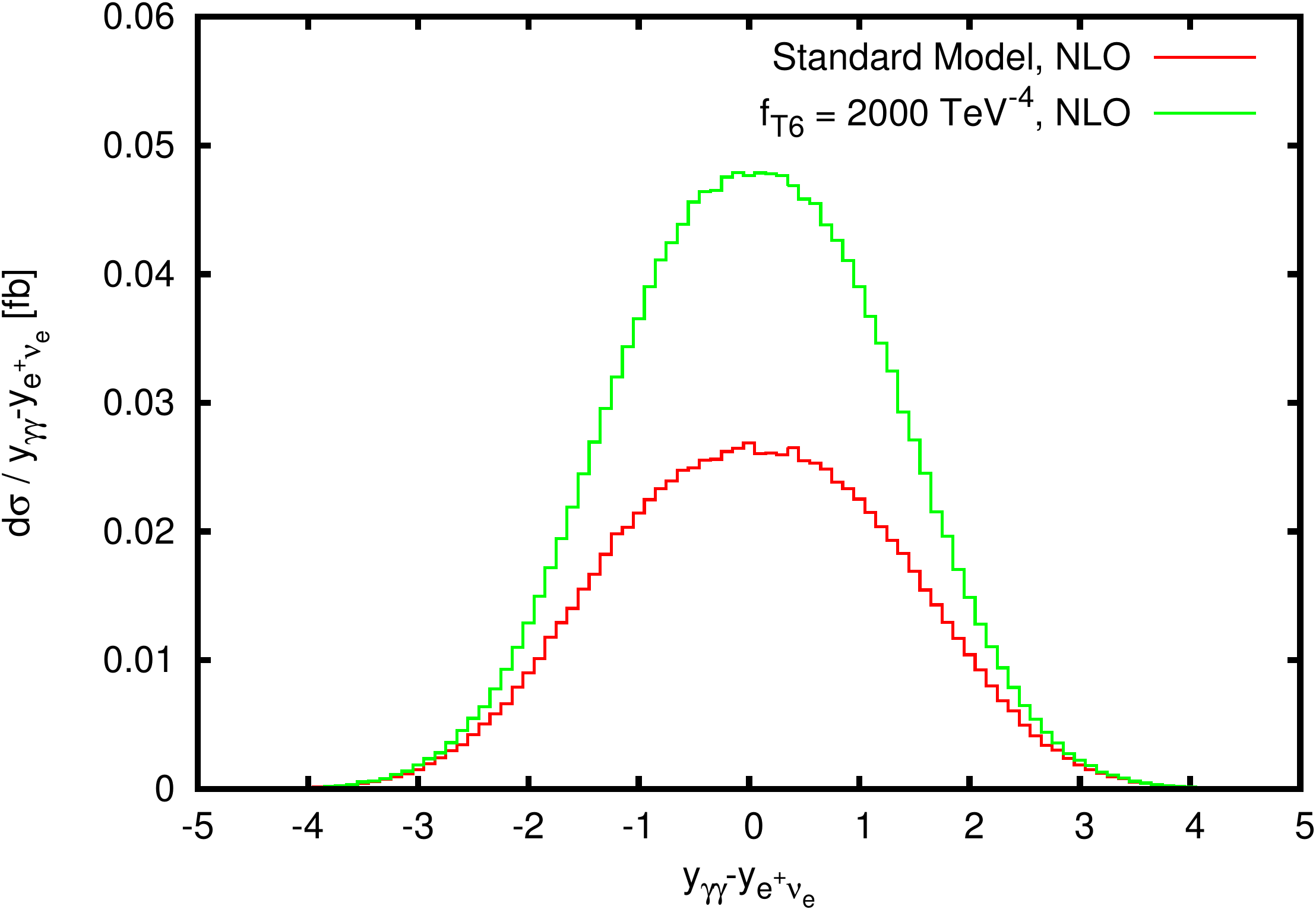}
\end{center}
\caption{Rapidity difference of the diphoton system and the
lepton-neutrino system for the SM and the anomalous coupling scenario.
\textit{Left:} LO distributions \textit{Right:} NLO distributions} 
\label{NLO:VVV:radzero}
\end{figure}

That this is indeed the case can be seen in Fig.~\ref{NLO:VVV:radzero}. 
Here we require additionally that the
transverse momentum of the harder photon exceeds 200 GeV and the
invariant mass of the lepton-neutrino system exceeds 75 GeV to suppress
radiation off the final-state lepton. The effect of the radiation zero
should be visible as a dip at zero in the rapidity difference between
the diphoton system and the lepton-neutrino system, which can be indeed
observed for the LO SM curve. In contrast the anomalous-coupling curve
shows no such behavior even at LO, and at NLO the dip is filled in both
cases.

Turning back to the right-hand plot of Fig.~\ref{NLO:VVV:pTA1}, the
ratio between anomalous-coupling and SM prediction decreases when going
from LO to NLO. This is due to the same effect, as part of the
additional contribution is caused by filling up the radiation zero,
which is no longer present at NLO because there already QCD effects have
caused this. Hence, for this process group, higher-order corrections
play an important role and cannot be neglected when determining the size
of or limits on anomalous quartic gauge couplings.

\subsubsection{Electroweak corrections to triboson processes}
\label{sec:tribosonew}

The first calculation of electroweak NLO corrections for a triboson
processes at hadron colliders has appeared only very recently. Hence, no
publicly available Monte Carlo implementation is available at the
present stage. For gauge boson pair production via vector-boson fusion
electroweak corrections no results exist in the literature at the
current stage.

In Ref.~\cite{Nhung:2013jta} the full NLO corrections to on-shell $WWZ$
production have been considered. Besides the QCD corrections already
calculated in Refs.~\cite{Hankele:2007sb,Binoth:2008kt}, additional
virtual electroweak diagrams with loops up to the pentagon level appear
as well as real-emission processes with an additional external photon.
There, processes with both photon radiation and initial-state photons
are taken into account. The latter appear when using PDFs with
photons~\cite{Martin:2004dh,Ball:2013hta}. Additionally, in this case
the photon-initiated contribution of $\gamma\gamma \rightarrow WWZ$ is
added at tree-level.
The electrweak corrections are typically quite small for integrated
cross sections, of about -2\%. They can, however, get significant in
differential distributions. For example, looking at the
transverse-momentum distribution of the $Z$ boson, at the 14 TeV LHC one
observes corrections of up to -30\% for transverse momenta of 1 TeV.
Thereby, the photon-initiated processes play an important role to partly
cancel large Sudakov virtual corrections.

\clearpage

\section{Comparison of predictions for multi-boson production with WHIZARD, VBFNLO and MADGRAPH}
\label{sec:comp}

Whenever more than one program is available to calculate the same
quantity, it is an important cross-check to ensure that the theory
predictions agree when choosing the same set of input parameters.
Therefore we compare the predictions for the three programs MadGraph5,
version 1.5.12, using the anomalous couplings implementation from
Ref.~\cite{Eboli:anom4}, \VBFNLO{} 2.7.0 beta 3 and WHIZARD
2.1.1. These three programs have been developed independently of each
other, so agreement provides a strong cross-check. As process we have
taken the same-sign $W$-pair vector-boson scattering process $pp
\rightarrow e^+ \nu_e \mu^+ \nu_\mu jj$. We calculate this process at
leading order for the LHC with a center-of-mass energy of 14 TeV using
the CTEQ6L1~\cite{Pumplin:2002vw} pdf set with a fixed factorization
scale $\mu=2 M_W$. No external bottom or top quarks are taken into
account. The SM electro-weak input parameters are set to
$M_W=80.398$~GeV, $M_Z=91.1876$~GeV, $M_H = 126$~GeV and $G_F =
1.16637 \cdot 10^{-5}$~GeV$^{-2}$ and the others fixed via
electro-weak tree-level relations. The widths of the bosons are
$\Gamma_W=2.097673$~GeV, $\Gamma_Z=2.508420$~GeV and
$\Gamma_H=4.277$~MeV.  All fermions are taken as massless. Cuts on the
final-state particles are as follows:
\begin{align}
p_{T,\ell} &> 20 \text{ GeV} & |\eta_\ell| &< 2.5 \nonumber\\ 
p_{T,j} &> 30 \text{ GeV} & |\eta_j| &< 4.5 \nonumber\\
|\Delta \eta_{jj}| &> 4 & M_{jj} &> 600 \text{ GeV} \ .
\label{eq:VBScomp:cuts}
\end{align}
The \VBFNLO{} program neglects any s-channel diagrams appearing, while for
the other two codes these are included as well. Their numerical impact
is, however, negligible due to the large invariant mass cut of the
two jets.
As anomalous quartic gauge couplings we take $f_{S,0} = f_{S,1} = \pm 10
\text{ TeV}^{-4}$ in the VBFNLO and MadGraph5 scheme, which corresponds
to $a_4 = \pm 4.59 \cdot 10^{-3}$ in WHIZARD. With these choices
unitarity would get violated at 1.2 TeV. Due to the more technical
nature of the comparison, and as there is no unitarization scheme
supported commonly between all three programs, we do not take this into
account further.

\begin{table}[b]
\begin{tabular}{l|c|c|c}
   & Standard Model & 
     $f_{S,0} = f_{S,1} = + 10 \text{ TeV}^{-4}$ &
     $f_{S,0} = f_{S,1} = - 10 \text{ TeV}^{-4}$ \\\hline
MadGraph5    & 1.3062(2) fb & 1.7918~(2) fb & 1.8295~(2) fb \\
\VBFNLO{}    & 1.3098(4) fb & 1.7932~(7) fb & 1.8310~(7) fb \\
WHIZARD      & 1.3094(8) fb & 1.7951(10) fb & 1.8325(12) fb 
\end{tabular}
\caption{Integrated cross section for the process $pp \rightarrow e^+
\nu_e \mu^+ \nu_\mu jj$ with the cuts defined in
Eq.~\ref{eq:VBScomp:cuts}. Results are given for all three programs for
both the SM and both signs in the anomalous coupling choice.}
\label{tab:VBScomp:cs}
\end{table}
In Table~\ref{tab:VBScomp:cs} we show results for the integrated
cross section for the SM and both signs of the anomalous coupling
choice. All three codes show a very good agreement with deviations of
only a few per mill. Between \VBFNLO{} and WHIZARD the level of
agreement is compatible with statistical fluctuations from Monte Carlo
integration, while the MadGraph5 result is slightly lower in all three
cases. 
To verify that this is not due to a mismatch in the input parameters, we
have compared the squared matrix element of the subprocess $uc
\rightarrow e^+ \nu_e \mu^+ \nu_\mu ds$, which has no s-channel
contributions, between MadGraph5 and \VBFNLO{} for 100 randomly chosen
phase-space points. Here we find excellent agreement at the sub-per mill
level between the two codes.

\begin{figure}
\includegraphics[width=0.48\textwidth]{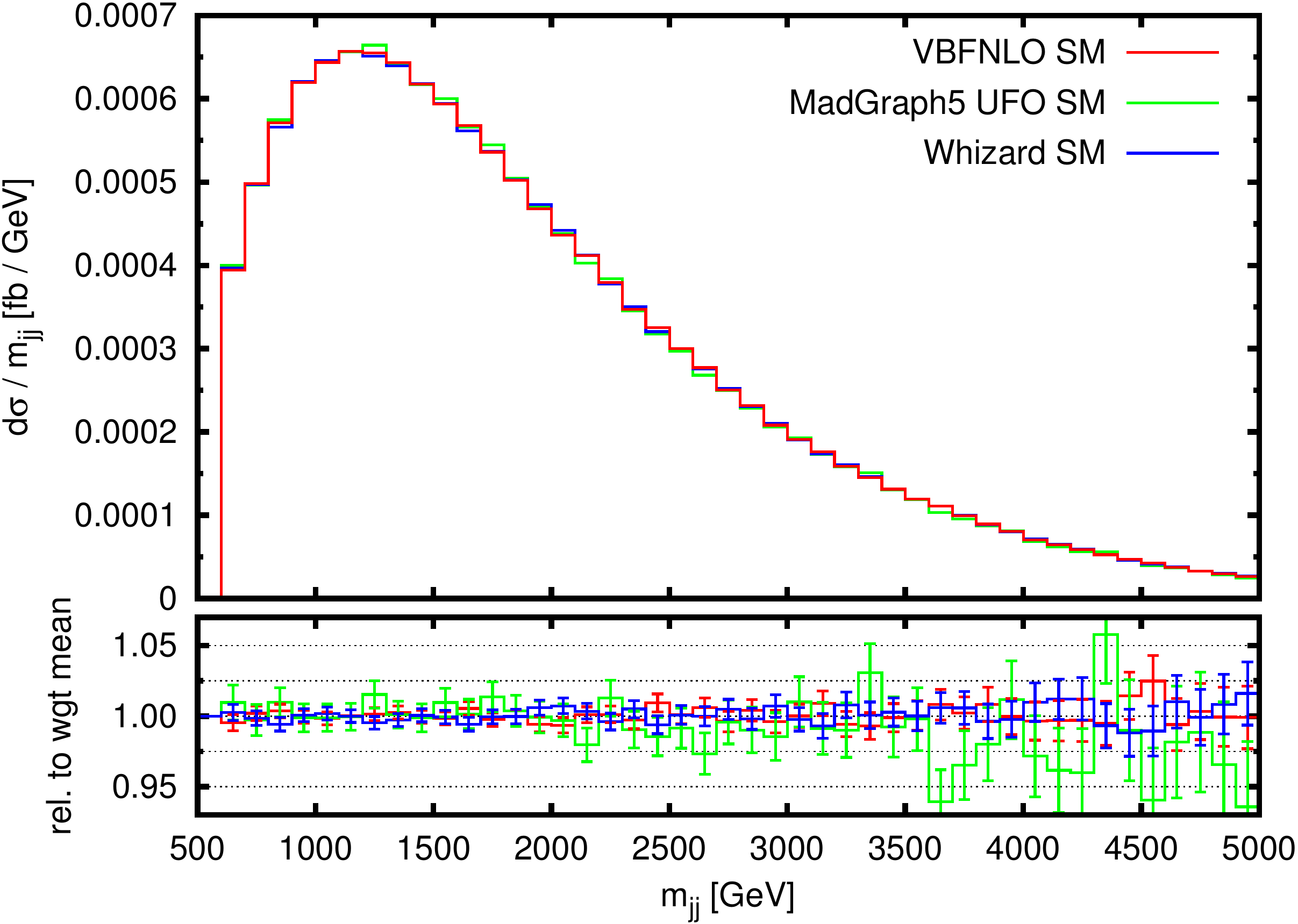}\quad
\includegraphics[width=0.48\textwidth]{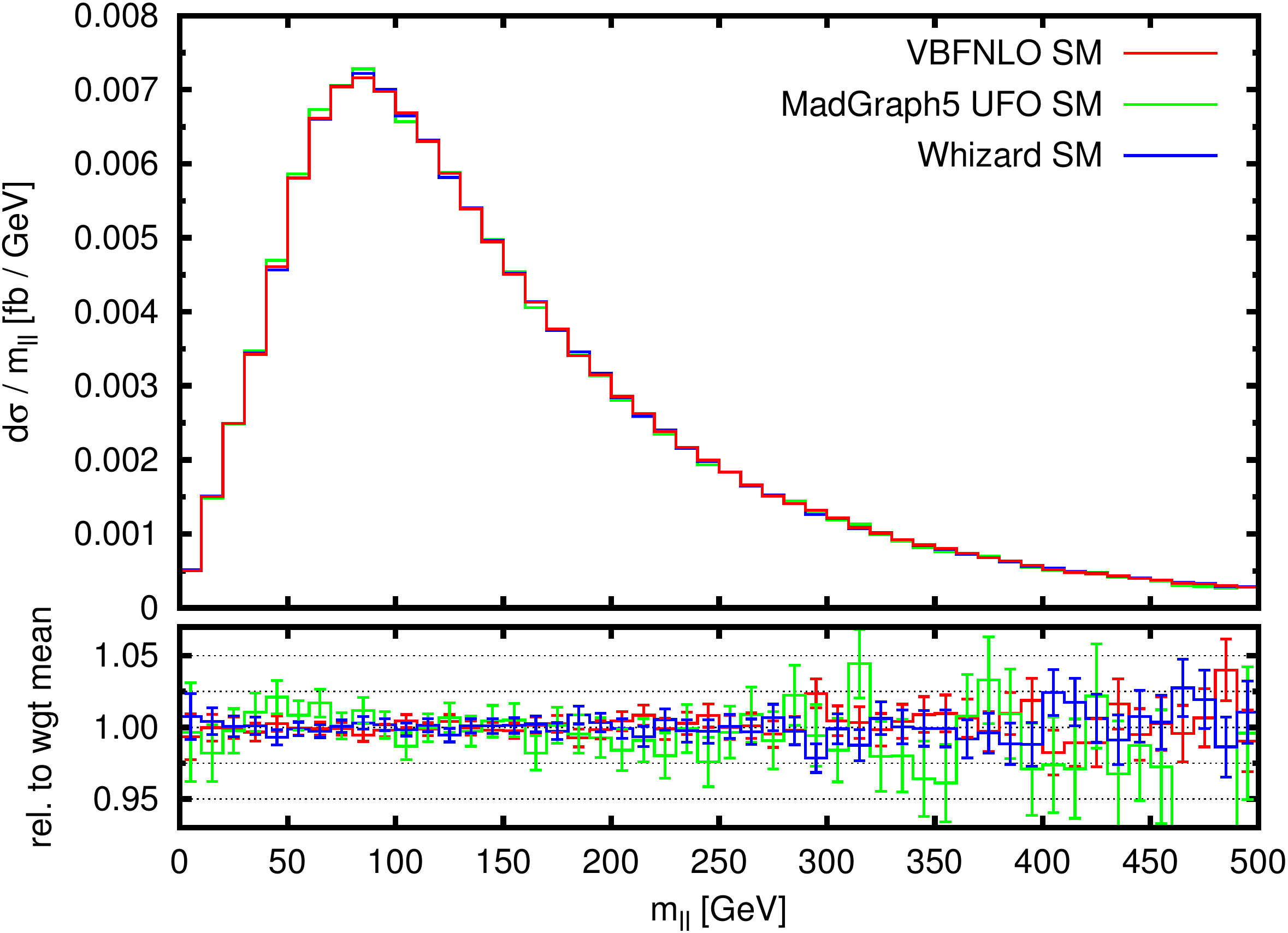}\\
\includegraphics[width=0.48\textwidth]{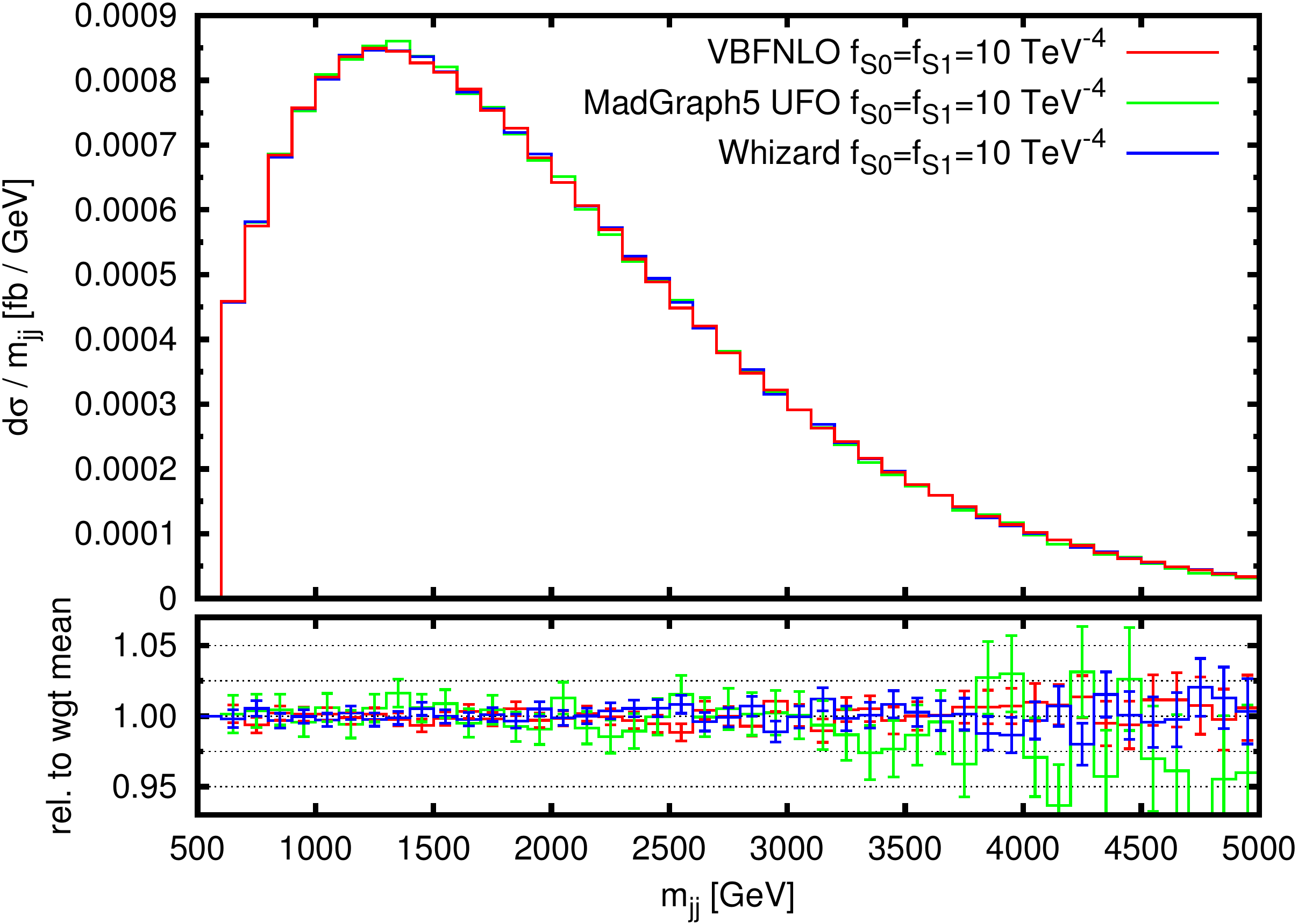}\quad
\includegraphics[width=0.48\textwidth]{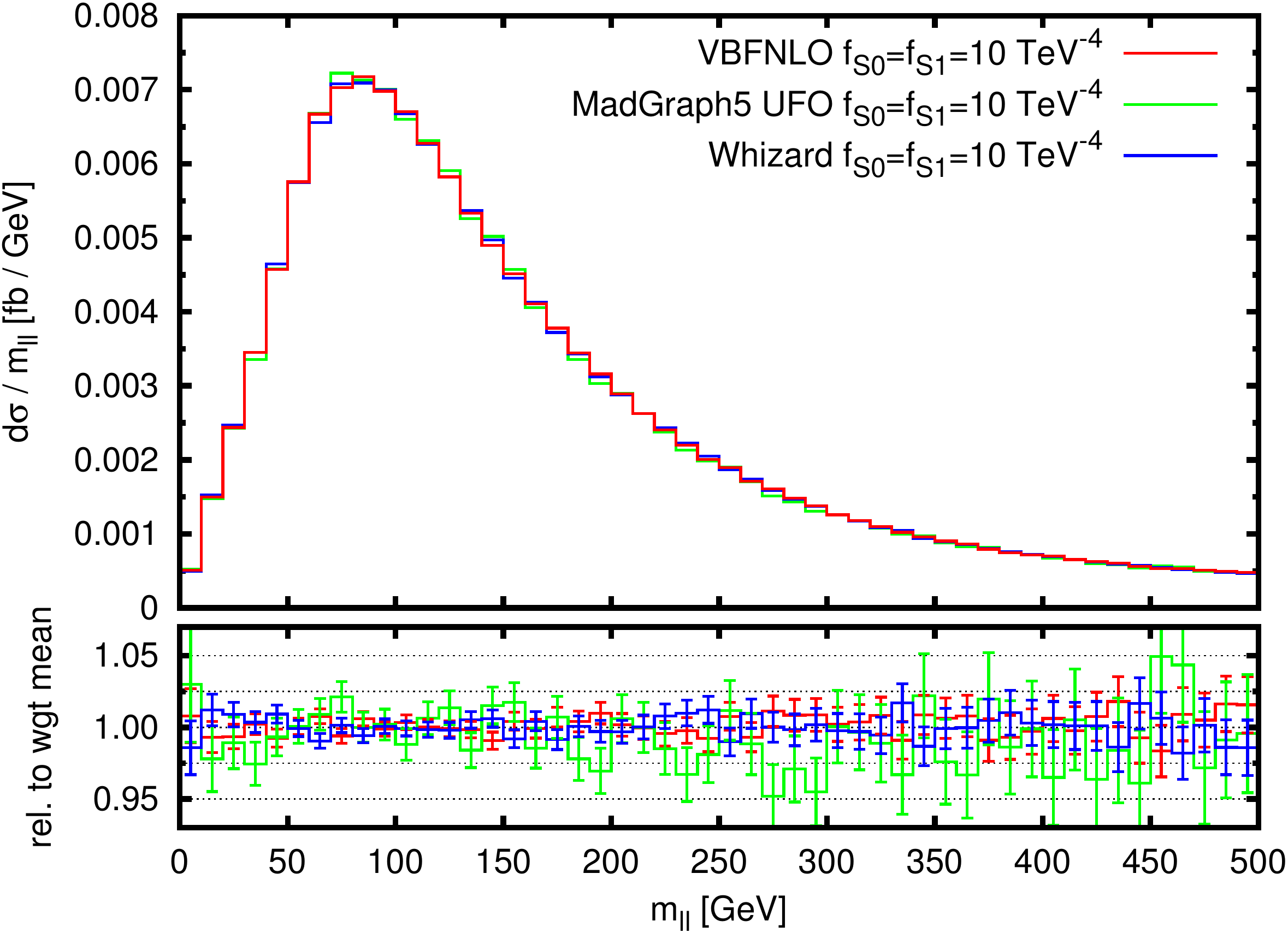}\\
\includegraphics[width=0.48\textwidth]{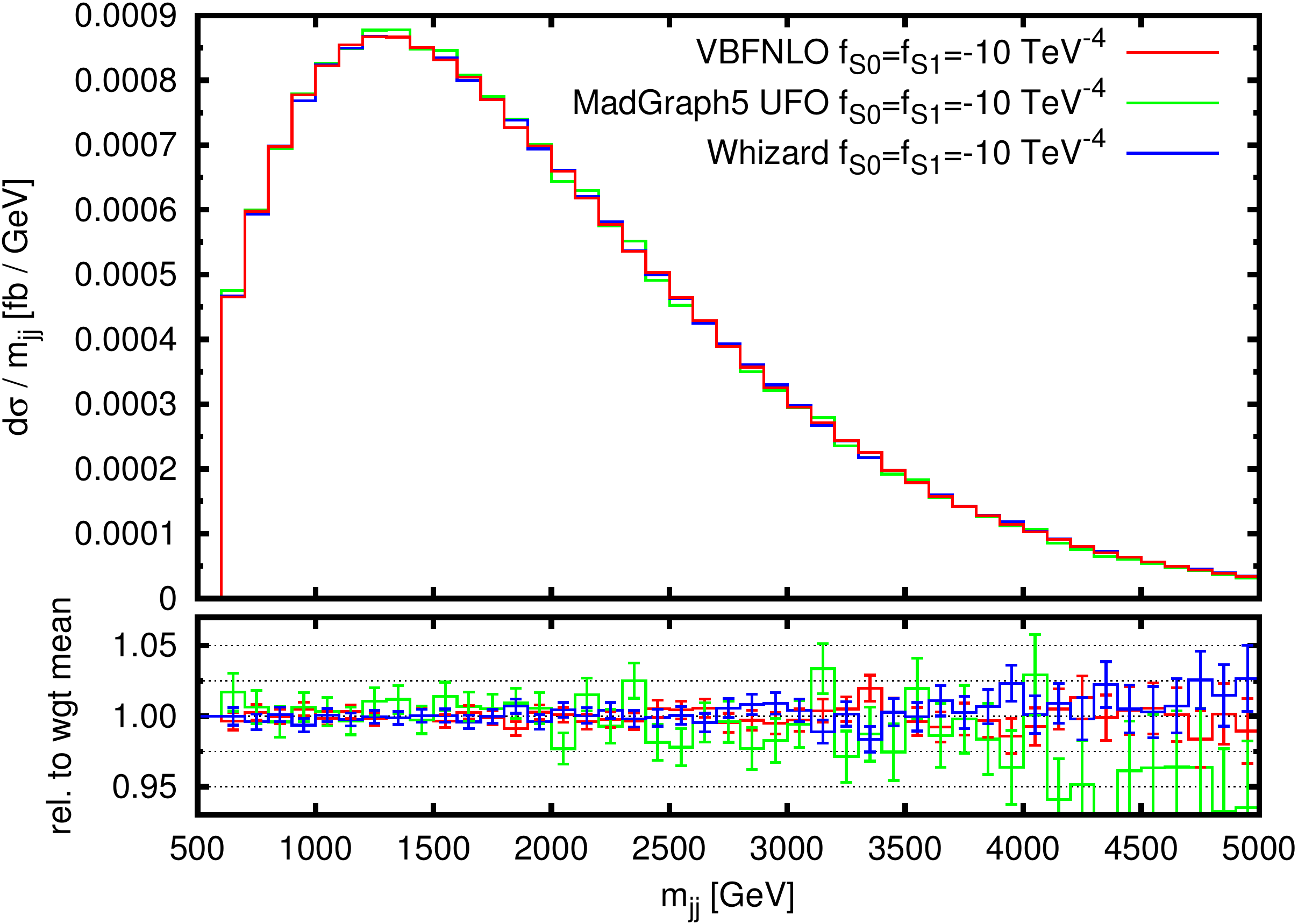}\quad
\includegraphics[width=0.48\textwidth]{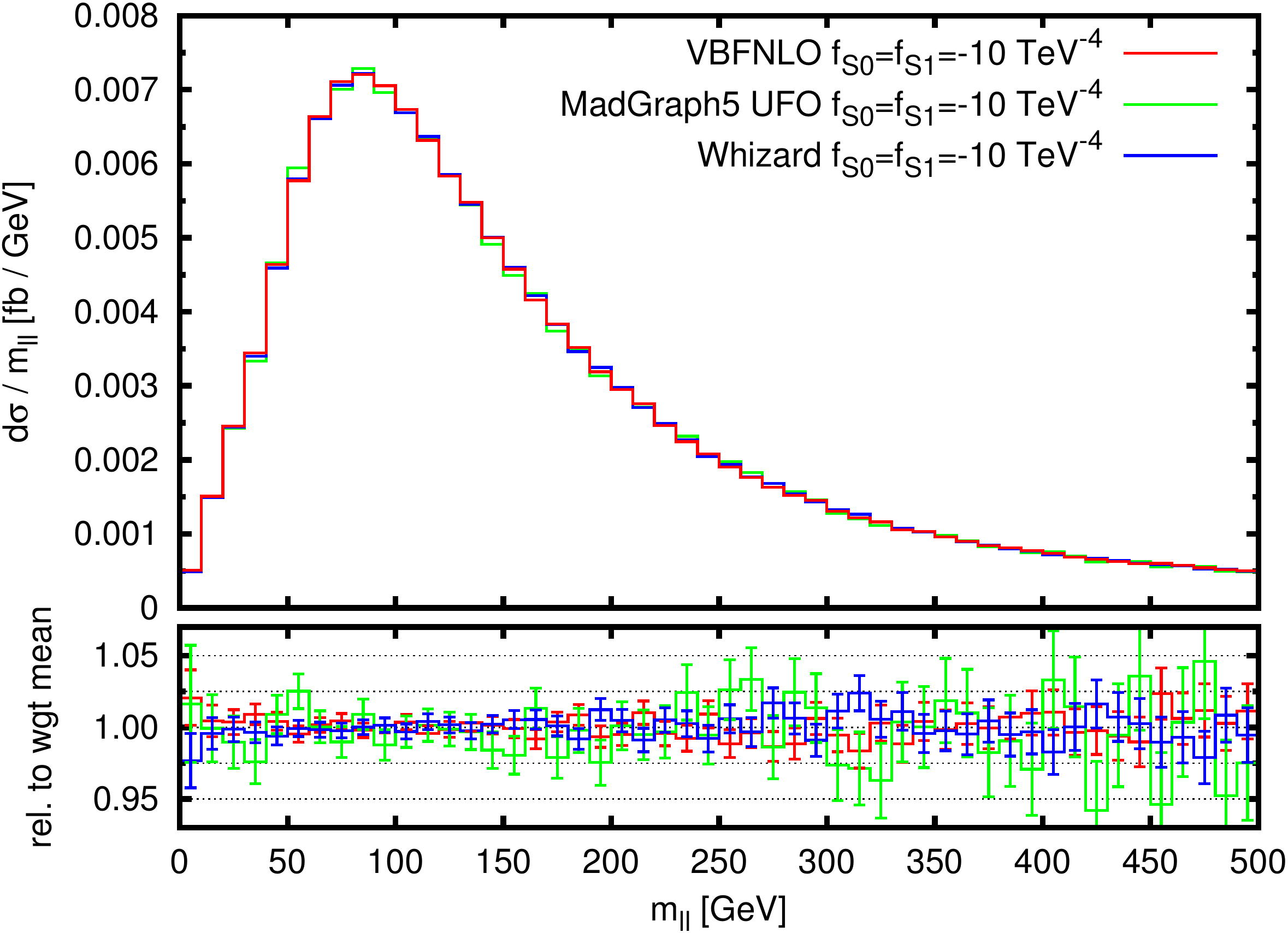}
\caption{Comparison of differential distributions between MadGraph5,
\VBFNLO{} and WHIZARD for the process $pp \rightarrow e^+
\nu_e \mu^+ \nu_\mu jj$.
\textit{Left:} Invariant mass of the two jets, 
\textit{Right:} invariant mass of the two charged leptons. 
\textit{Top:} SM,
\textit{Middle:} positive anomalous coupling,
\textit{Bottom:} negative anomalous coupling. The upper part of each
plot shows the differential cross section, while the lower part shows the
ratio of each code to the weighted mean of all three codes.}
\label{fig:VBScomp:dist}
\end{figure}
In Fig.~\ref{fig:VBScomp:dist} we then compare differential distributions
between MadGraph5, \VBFNLO{} and WHIZARD. Each code has been asked to
generate 1 million unweighted events, which form the input of each plot. 
The left column shows the invariant mass of the two jets, while in the
right column the invariant mass of the two charged leptons is plotted.
The top row presents each distribution for the SM, while in the middle
and bottom row results for positive and negative anomalous couplings are
shown, respectively. The lower part of each plot shows the differential
cross section of each code compared to the weighted average of all three
codes. 

Similar to the integrated results, there is good agreement between
all three codes. Deviations from the weighted average are compatible
with those from finite event statistics, indicated by the error bars. No
systematic shifts are visible, although MadGraph5 does tend to favor
slightly smaller differential cross sections in the large $m_{jj}$
range.
Comparing the two distributions between the SM and the two anomalous
coupling scenarios, we see that the shape of the $m_{jj}$ distribution
hardly changes. The situation is different for the $m_{\ell\ell}$
distribution. Here we observe that for low invariant masses the
distribution receives no additional contribution. This can for example
seen when looking at the bin with the largest differential cross
section, whose height stays approximately the same. On the other hand,
for larger invariant masses a significant increase of the differential
cross section happens. Such a behavior is not surprising, as the
invariant mass of the two leptons is directly related to the invariant
mass of the $WW$ system, and therefore one expects that the effects on
anomalous couplings become larger for larger values, while no such link
exists for the invariant mass of the two jets.

To further corroborate the agreement in the implementation of anomalous
quartic gauge couplings, we have performed an additional cross-check
between MadGraph5 and \VBFNLO{} calculating the triboson process $pp
\rightarrow e^+ \nu_e \gamma\gamma$. As anomalous quartic gauge coupling
we choose the operator $M2$ with numerical value 
$f_{M2} = 8187 \text{ TeV}^{-4}$ in the \VBFNLO{} and 
$f_{M2} = -250 \text{ TeV}^{-4}$ in the MadGraph5 convention.
The integrated cross sections are $1.8012(8)$ fb and $1.8172(5)$ fb in
the SM case and $4.2482(19)$ fb and $4.2660(13)$ fb including the anomalous
quartic gauge coupling, where the first value in both cases refers to
MadGraph5 and the second one to \VBFNLO{}, respectively.
While some difference exceeding the statistical errors from Monte Carlo
integration is also present here, the agreement is at the sub-percent level for
both scenarios and hence good.

\begin{figure}
\includegraphics[width=0.48\textwidth]{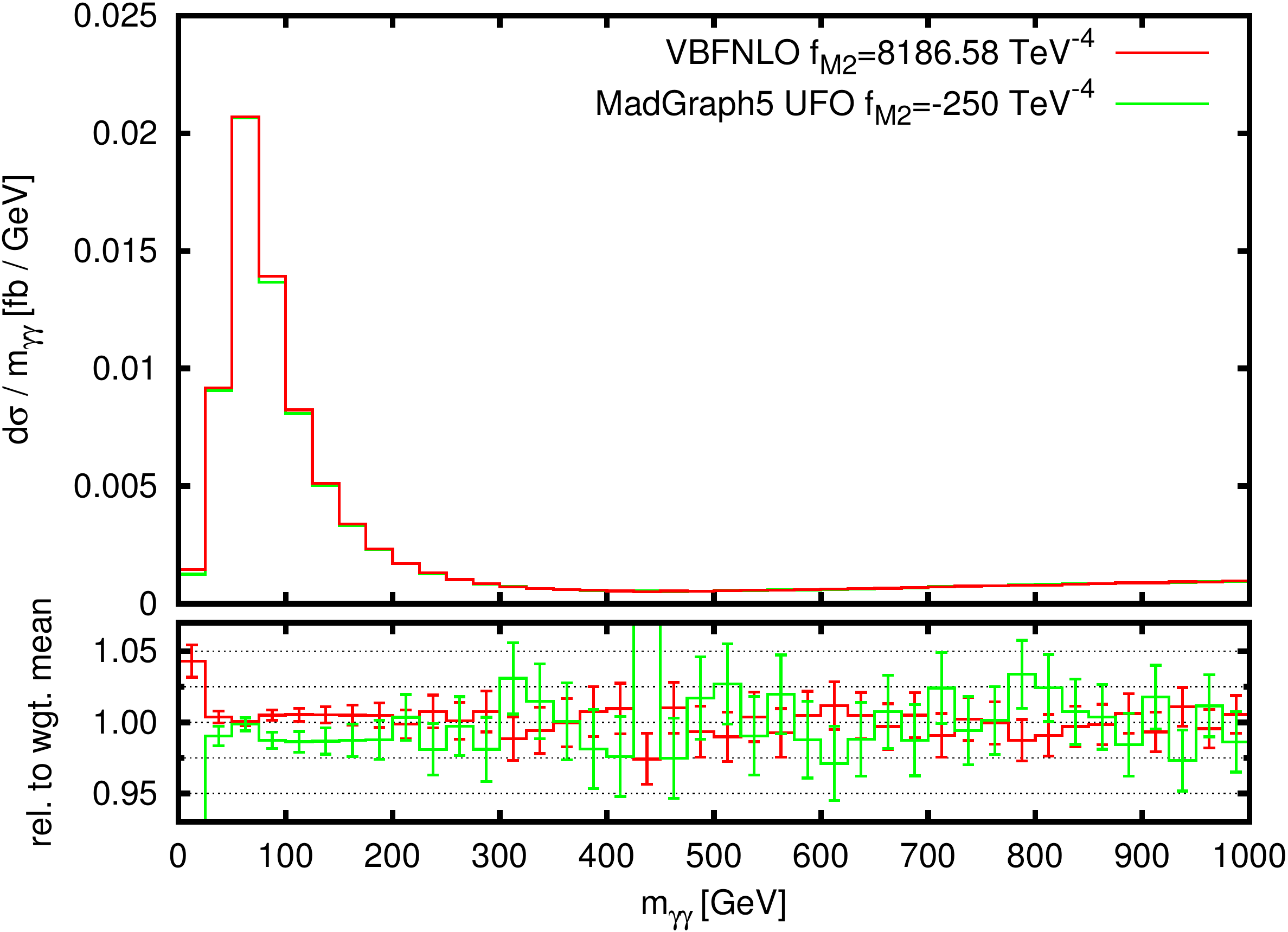}\quad
\includegraphics[width=0.48\textwidth]{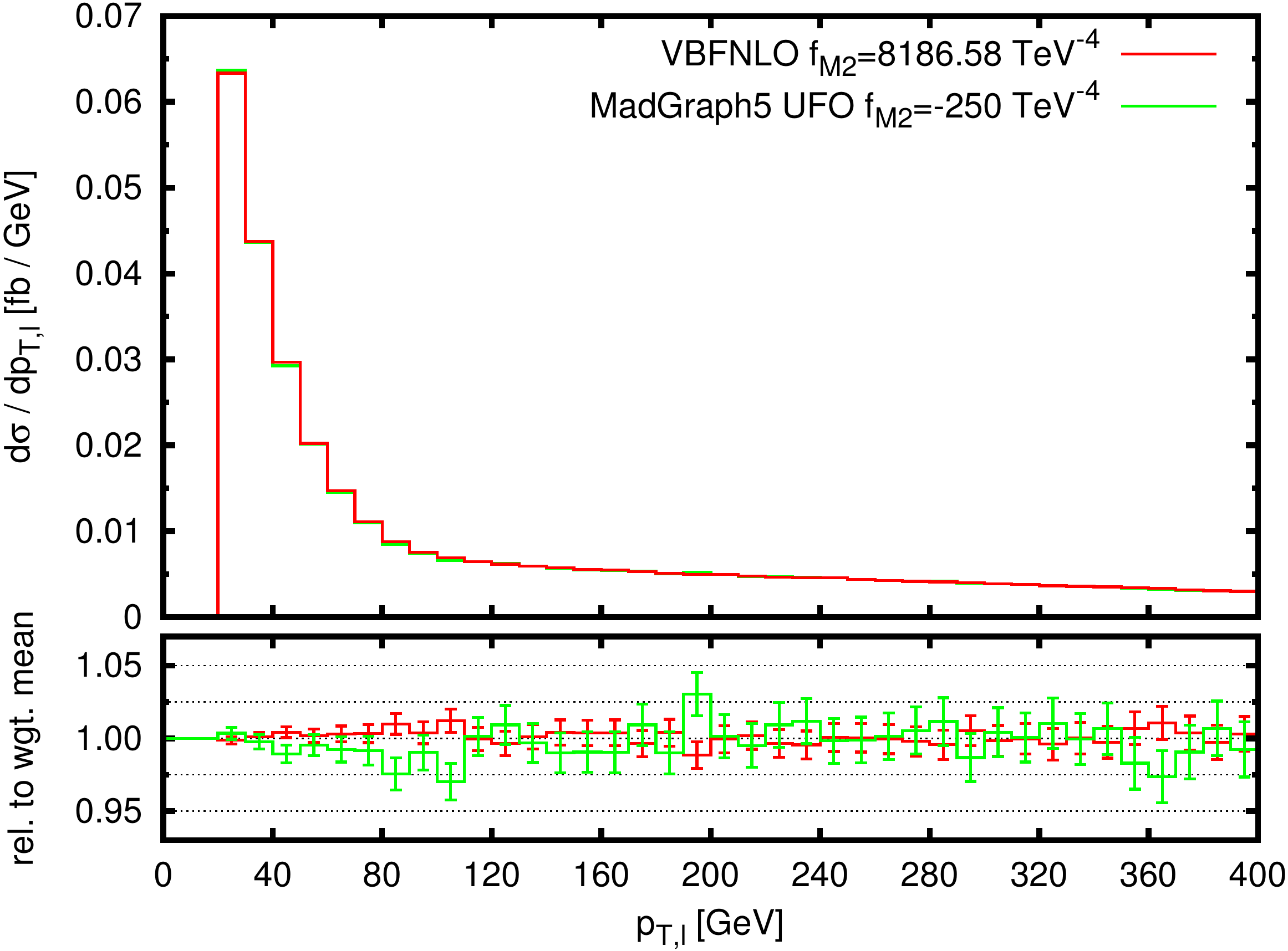}\\
\caption{Comparison of differential distributions between MadGraph5 and
\VBFNLO{} for the process $pp \rightarrow e^+ \nu_e \gamma\gamma$ using
the anomalous quartic gauge coupling operator $M2$.
\textit{Left:} Invariant mass of the photon pair, 
\textit{Right:} transverse momentum of the lepton. 
}
\label{fig:Triboson:dist}
\end{figure}
In Fig.~\ref{fig:Triboson:dist} we show the differential cross section
for the invariant mass of the photon pair on the left-hand side and for
the transverse momentum of the lepton on the right-hand side. In both
cases we observe a reasonable agreement between the two codes within
statistical errors. We have also checked several other distributions and
do not see any deviations that would be incompatible with an explanation
by statistical effects.

\clearpage

\section{Summary}
\label{sec:summary}

In this Snowmass 2013 white paper we presented an overview of the
theory of electroweak non-standard interactions and of publicly
available Monte Carlo tools that provide predictions for electroweak
vector boson pair and triple production as well as vector boson
scattering at the LHC, including non-standard EW interactions. We
reviewed the role of higher-order corrections in the study of
non-standard EW couplings in these processes, using \VBFNLO{} and a
\POWHEGBOX{} implementation of higher-order QCD corrections to $WWjj$
production. We performed a tuned comparison of predictions obtained
with {\textsc MadGraph5}, \VBFNLO{}, and WHIZARD for a number of
relevant observables at leading order QCD and including
higher-dimension operators in EFT, and found good agreement.

\acknowledgments
C.~D. is supported in part by the U.~S. Department of Energy under Contract No. DE-FG02-13ER42001.
B.~F. and M.~R. would like to thank D.~Zeppenfeld for many helpful and
stimulating discussions.  B.~F. and M.~R. acknowledge support by the
BMBF under Grant No. 05H09VKG (``Verbundprojekt HEP-Theorie'').
B.~J. would like to thank the Research Center Elementary Forces and
Mathematical Foundations (EMG) of the Johannes Gutenberg University
Mainz. 
O.~M. is a fellow of the Belgian American Educational Foundation and a chercheur logistique post-doctoral from F.R.S-FNRS.
His work is partially supported by the IISN MadGraph convention 4.4511.10.
The work of D.~W. is supported in part by the U.S.~National
Science Foundation under grant no. PHY-1118138.

\clearpage

\end{document}